\documentclass[twocolumn]{aastex62}

\usepackage{multirow}
\usepackage{amsmath}
\usepackage{longtable}
\usepackage{color}

\newcommand{\vlsr}     {V_\mathrm{lsr}}
\newcommand{\vsys}     {V_\mathrm{sys}}
\newcommand{\vout}     {V_\mathrm{out}}
\newcommand{\vrot}     {V_\mathrm{rot}}
\newcommand{\pa} {\mathrm{P.A.}}
\newcommand{\vcb}     {v_\mathrm{CB}}
\newcommand{\rcb}     {r_\mathrm{CB}}
\newcommand{\rout}     {r_\mathrm{out}}
\newcommand{\tex}     {T_\mathrm{ex}}

\newcommand{\h} {^{\mathrm{h}}}
\newcommand{\m} {^{\mathrm{m}}}
\newcommand{\s} {^{\mathrm{s}}}

\newcommand{\mJybeam}  {\mbox{mJy}~\mbox{beam}^{-1}}

\newcommand{\kms}	{\mbox{km s}^{-1}}
\newcommand{\pc}	{\mbox{pc}}
\newcommand{\yr}	{{\rm yr}}
\newcommand{\K}	{{\rm K}}

\newcommand{\au} {\mbox{au}}

\shorttitle{Ordered Envelope-disk Transition in G339.88-1.26}
\shortauthors{Zhang et al.}

\begin{document}

\title{An Ordered Envelope-disk Transition in the Massive Protostellar Source G339.88-1.26}

\author{Yichen Zhang}
\affiliation{Star and Planet Formation Laboratory, RIKEN Cluster for Pioneering Research, Wako, Saitama 351-0198, Japan; yichen.zhang@riken.jp}

\author{Jonathan C. Tan}
\affiliation{Department of Space, Earth \& Environment, Chalmers University of Technology, SE-412 96 Gothenburg, Sweden}
\affiliation{Department of Astronomy, University of Virginia, Charlottesville, VA 22904-4325, USA}

\author{Nami Sakai}
\affiliation{Star and Planet Formation Laboratory, RIKEN Cluster for Pioneering Research, Wako, Saitama 351-0198, Japan; yichen.zhang@riken.jp}

\author{Kei E. I. Tanaka}
\affiliation{Department of Earth and Space Science, Osaka University, Toyonaka, Osaka 560-0043, Japan}
\affiliation{ALMA Project, National Astronomical Observatory of Japan, Mitaka, Tokyo 181-8588, Japan}

\author{James M. De Buizer} 
\affiliation{SOFIA-USRA, NASA Ames Research Center, MS 232-12, Moffett Field, CA 94035, USA}

\author{Mengyao Liu}
\affiliation{Department of Astronomy, University of Virginia, Charlottesville, VA 22904-4325, USA}

\author{Maria T. Beltr\'an} 
\affiliation{INAF -- Osservatorio Astrofisico di Arcetri, Largo E. Fermi 5, 50125 Firenze, Italy}

\author{Kaitlin Kratter}
\affiliation{Department of Astronomy and Steward Observatory, University of Arizona, 933 N Cherry Ave, Tucson, AZ, 85721, USA}

\author{Diego Mardones}
\affiliation{Departamento de Astronom\'ia, Universidad de Chile, Casilla 36-D, Santiago, Chile}

\author{Guido Garay}
\affiliation{Departamento de Astronom\'ia, Universidad de Chile, Casilla 36-D, Santiago, Chile}

\begin{abstract}
We report molecular line observations of the massive protostellar
source G339.88-1.26 with the Atacama Large Millimeter/Submillimeter
Array.  The observations reveal a highly collimated SiO jet extending
from the 1.3~mm continuum source, which connects to a slightly wider
but still highly collimated CO outflow.  Rotational features
perpendicular to the outflow axis are detected in various molecular
emissions, including SiO, SO$_2$, H$_2$S, CH$_3$OH, and H$_2$CO
emissions.  Based on their spatial distributions and kinematics, we
find that they trace different parts of the envelope-disk system. 
The SiO emission traces the disk and inner envelope in addition to the jet.
The CH$_3$OH and H$_2$CO emissions mostly trace
the infalling-rotating envelope, and are enhanced around the transition region between
envelope and disk, i.e., the centrifugal barrier. The SO$_2$
and H$_2$S emissions are enhanced around the centrifugal barrier, and also trace
the outer part of the disk.  Envelope kinematics are consistent with rotating-infalling
motion, while those of the disk are consistent with Keplerian
rotation.  The radius and velocity of the centrifugal barrier are
estimated to be about 530 au and $6~\kms$, leading to a central mass of
about $11\:M_\odot$, consistent with estimates based on spectral
energy distribution fitting.  These results indicate that an ordered
transition from an infalling-rotating envelope to a Keplerian disk
through a centrifugal barrier, accompanied by changes of types of molecular line emissions, 
is a valid description of this massive protostellar
source.  This implies that at least some massive stars form in a
similar way as low-mass stars via Core Accretion.
\end{abstract}

\keywords{ISM: individual objects (G339.88-1.26), molecules, 
kinematics and dynamics, jets and outflows --- stars: formation, massive}

\section{Introduction}
\label{sec:intro}

Massive stars impact many areas of astrophysics, yet there is little
consensus on how they form. One of the key questions is whether
massive protostars accrete through rotationally supported, i.e.,
Keplerian or near Keplerian, disks, as have been seen around low-mass
protostars.  High angular resolution observations, especially of gas
kinematics with molecular lines, have provided a handful of candidates
of Keplerian disks around massive protostars
(\citealt[]{Beltran16}). Most of these disks are around B-type
protostars (up to about $15\:M_\odot$; e.g.,
\citealt[]{Sanchez13,Beltran14,Ginsburg18}), with just a few examples
reported from around O-type massive stars (e.g.,
\citealt[]{Johnston15,Zapata15,Ilee16,Cesaroni17,Maud18}).  More often,
especially around O-type protostars, massive ($\sim100\:M_\odot$)
rotating ``toroids'' of radii of $10^3-10^4\:\au$ are found. It is
unclear whether such toroids are feeding smaller Keplerian disks at
their centers. Searching for Keplerian disks around massive protostars
is challenging due to the far distances and embedded, crowded
environments.  Often the kinematics of putative Keplerian disks and
outer infalling-rotating envelopes are not easily distinguished.

In low-mass star formation, the transition from an infalling-rotating
envelope to a rotationally supported disk has been much better observed. 
We note that in this paper, by the term ``disk'', we only refer to a
rotationally supported disk, i.e., where rotation is the dominant form
of support against infall,
and we also do not distinguish an infalling-rotating envelope with an
infalling-rotating ``pseudo disk'' (also see discussions in
\S\ref{sec:caveats}).  As the material in the core collapses, the
infalling motion is gradually converted to rotation. The innermost
radius that such material can reach without losing angular momentum is
the centrifugal barrier (\citealt[]{Sakai14}; see also
\citealt[]{Stahler94}), inside of which a rotationally supported disk
is expected. The centrifugal barrier is accompanied not only with a
change of kinematics from infalling-rotation to Keplerian rotation,
but also a change of chemical compositions due to the accretion shock
and different temperature and density conditions in the disk and
envelope (e.g., \citealt[]{Sakai14b,Oya15,Oya16,Oya17}). For example,
in the low-mass sources studied by these authors, the
infalling-rotating envelopes outside of the centrifugal barriers are
often traced by molecules such as CCH, c-C$_3$H$_2$, CS, and OCS.  The
accretion shocks associated with the centrifugal barriers are often
highlighted by volatile species such as SO and saturated organic
species.  The Keplerian disks can be traced by molecules such as
C$^{18}$O and H$_2$CO.  The observations of different molecules help
to disentangle the disk and envelope, and highlight the transition
region, which provide a powerful diagnostic tool to understand the
whole picture of disk formation.

There are two main theories for massive star formation: Core
Accretion, which is a scaled-up version of low-mass star formation
(e.g., \citealt[]{MT03}); and Competitive Accretion
(e.g., \citealt[]{Bonnell01,Wang10}), in which stars chaotically gain
their mass via the global collapse of the clump without passing
through the massive core phase (see \citealt[]{Tan14} for a review).  
In the Core Accretion scenario, 
a transition from relatively ordered rotating infall of a massive core at scales
of $\sim10^3\:\au$ to a rotationally supported disk on scales of
several $\times 10^2\:\au$ or even smaller is expected,
and these different components may be highlighted by emissions of
different molecules, similar to the case of low-mass star formation,
though the particular molecules tracing the various components may be different
from the case of low-mass star formation,
due to different temperature, density and shock conditions.  
In the Competitive Accretion model, disks are also expected, but are likely to be much
smaller due to close protostellar interactions. In addition, since the
collapse is more disordered, a simple envelope-disk transition,
companied by clear change of kinematic and chemical patterns,
is not expected. 
Therefore searching for Keplerian disks and understanding how and 
where the envelope-disk transition happens by using multiple molecular lines
is important to test different theories of massive star formation.

Recent observations have shown indication of a centrifugal barrier 
between the envelope and disk around the
massive protostar G328.2551-0.5321 from the enhancement of
high-excitation CH$_3$OH emissions (\citealt[]{Csengeri18}).  However,
so far, there are no simultaneous confirmation of centrifugal barriers
around massive protostars using both kinematic and chemical features.
In this paper, we report observations of Atacama Large
Millimeter/Submillimeter Array (ALMA) on the massive protostellar
source G339.88-1.26, revealing the transition from an
infalling-rotating envelope to a Keplerian disk, identified by both
kinematic and chemical patterns.

\section{The Target and Observations}

\subsection{The Target G339.88-1.26}
\label{sec:target}

\begin{table*} 
\scriptsize
\begin{center}
\caption{Parameters of the Observed Lines\footnote{Line information taken from the 
CDMS database (\citealt[]{Muller05})} \label{tab:lines}}
\begin{tabular}{lccccccc}
\hline
\hline
Molecule & Transition & Frequency & $E_u/k$ & $S\mu^2$ & Velocity Resolution & Synthesized Beam & Channel rms\\
 & & (GHz) & (K) & (D$^2$) & ($\kms$) & & ($\mJybeam$)\\
\hline
$^{12}$CO & $2-1$ & 230.5380000 & 16.6 & 0.0242 & 0.63 & $0.27\arcsec\times 0.20\arcsec$  ($\pa=29.3^\circ$) & 4.1 \\
C$^{18}$O & $2-1$ & 219.5603541 & 15.8 & 0.0244 & 1.67 & $0.29\arcsec\times 0.22\arcsec$  ($\pa=26.5^\circ$) & 5.7 \\
SiO & $5-4$ & 217.1049190 & 31.3 & 48.0 & 0.67 & $0.30\arcsec\times 0.22\arcsec$  ($\pa=28.5^\circ$) & 2.9 \\
CH$_{3}$OH & $4_{2,2}-3_{1,2}$; E & 218.4400630 & 45.5 & 13.9 & 1.68 & $0.29\arcsec\times 0.21\arcsec$  ($\pa=28.2^\circ$) & 5.6 \\
H$_{2}$CO & $3_{2,1} - 2_{2,0}$ & 218.7600660 & 68.1 & 9.06 & 1.67 & $0.29\arcsec\times 0.21\arcsec$  ($\pa=28.2^\circ$) & 5.1 \\
SO$_{2}$ &  $22_{2,20}-22_{1,21}$ & 216.6433035 & 248 & 35.3 & 0.68 & $0.29\arcsec\times 0.22\arcsec$  ($\pa=28.5^\circ$) & 3.1 \\
H$_{2}$S & $2_{2,0} - 2_{1,1}$ & 216.7104365 & 84.0 & 2.06 & 0.68 & $0.29\arcsec\times 0.22\arcsec$  ($\pa=28.5^\circ$) & 3.1 \\
\hline
\end{tabular}
\end{center}
\end{table*}

Our target G339.88-1.26 (a.k.a. IRAS 16484-4603; hereafter G339) is a
massive protostellar source at a distance of $d = 2.1\:$kpc
(\citealt[]{Krishnan15}).  Interferometric radio continuum
observations revealed an elongated structure of $\sim 10\arcsec$ with
a position angle of $45^\circ$ east of north
(\citealt[]{Ellingsen96,Purser16}), which is believed to be tracing an
ionized jet.  6.7~GHz CH$_3$OH maser spots are found to be linearly
distributed in both space (within a scale of $\sim 1\arcsec$) and
velocity, with the spatial distribution approximately perpendicular to
the radio jet (\citealt[]{Ellingsen96}), which led to an explanation
that the masers are tracing a rotating disk.  Mid-Infrared (MIR)
observations at 10 and 18 $\mu$m revealed emissions on a scale of
$~4\arcsec$, elongated in about the east-west direction, which is
resolved into three peaks (1A, 1B, and 1C; \citealt[]{Debuizer02}).
\citet[]{Debuizer02} further argued that there are two stellar or
protostellar sources present.  One source, which corresponds to the
MIR emission peak 1B, is an embedded high-mass source driving the
radio jet.  Another source (about $1\arcsec$ to the west of 1B, and
not corresponding to any MIR emission peak) is a massive star slightly
in the foreground and less obscured, which is creating an extended HII
region (however see \S\ref{sec:multiwave}). In such a scenario, the masers are believed
not to trace the accretion disk, but are generated by shocks
associated with the embedded source.  Such an origin of the maser
emissions is also supported by the polarization observations of the
CH$_3$OH masers (\citealt[]{Dodson08}).  Based on spectral energy
distribution (SED) fitting using models of single massive protostars
(\citealt[]{ZT18}), the total luminosity of the source is estimated to
be about $(4-6)\times 10^4\:L_\odot$ and the protostellar mass is
estimated to be about $12-16\:M_\odot$ (\citealt[]{Liu19}).

\subsection{Observations}
\label{sec:obs}

The observations were carried out with ALMA in Band 6,
covering frequency range from 216 to 232 GHz, on April 4, 2016
with the C36-3 configuration and on September 15, 2016 with the C36-6
configuration.  The total integration time is 3.5 and 6.6 minutes in
the two configurations.  36 antennas were used in both
configurations. The baselines ranged from 15~m to 462~m in the C36-3
configuration, and from 27~m to 3.1~km in the C36-6 configuration.
J1427-4206 was used for bandpass calibration, J1617-5848 and Titan
were used for flux calibration, and J1636-4102 and J1706-4600 were
used as phase calibrators.  The source was observed with single
pointings, and the primary beam size (half power beam width) was
$22.9\arcsec$.  

The data were calibrated and imaged in CASA
(\citealt{McMullin07}).  
After pipeline calibration, we performed self-calibration
using the continuum data obtained from a 2~GHz-wide spectral window
with line-free bandwidth of about 1.6~GHz. 
We first performed two phase-only self-calibration iterations with
solution intervals of 30s and 6s, and then
one iteration of amplitude self-calibration with the solution interval
equal to the scan interval.
We applied the self-calibration phase and amplitude solutions
to the other spectral windows.
The peak signal-to-noise ratio of the continuum image is increased
by a factor of 2 by the self-calibration.
The data of two configurations were then combined after the
calibration and self-calibration of the individual data sets.
The resultant largest recoverable scale is about $11\arcsec$.  
To image the data, the CASA task clean was used, 
using robust weighting with the robust parameter of 0.5. 
For the continuum imaging, in addition to the 2~GHz-wide spectral window
(line-free bandwidth of 1.6~GHz),
we also combined the line-free
channels of other spectral windows,
making the total continuum bandwidth of about 2.3~GHz.  The synthesized
beam of the 1.3 mm continuum data is $0.29\arcsec\times 0.21\arcsec$ with
$\pa=28.2^\circ$.  The continuum peak position is derived to be
$(\alpha_{2000},\delta_{2000})=(16\h52\m04\s.663$,
$-46^\circ08\arcmin33\arcsec.88)$.  In this paper, we only focus on
several main lines that have been detected by our observation.  The
parameters of these lines are summarized in Table \ref{tab:lines}.  We
adopt a systemic velocity of the source of $\vsys=-33\:\kms$ based on
previous single-dish observations (\citealt[]{Debuizer09}).

\section{Distribution}
\label{sec:distribution}

\subsection{1.3 mm Continuum Emission}
\label{sec:continuum}

\begin{figure}
\begin{center}
\includegraphics[width=\columnwidth]{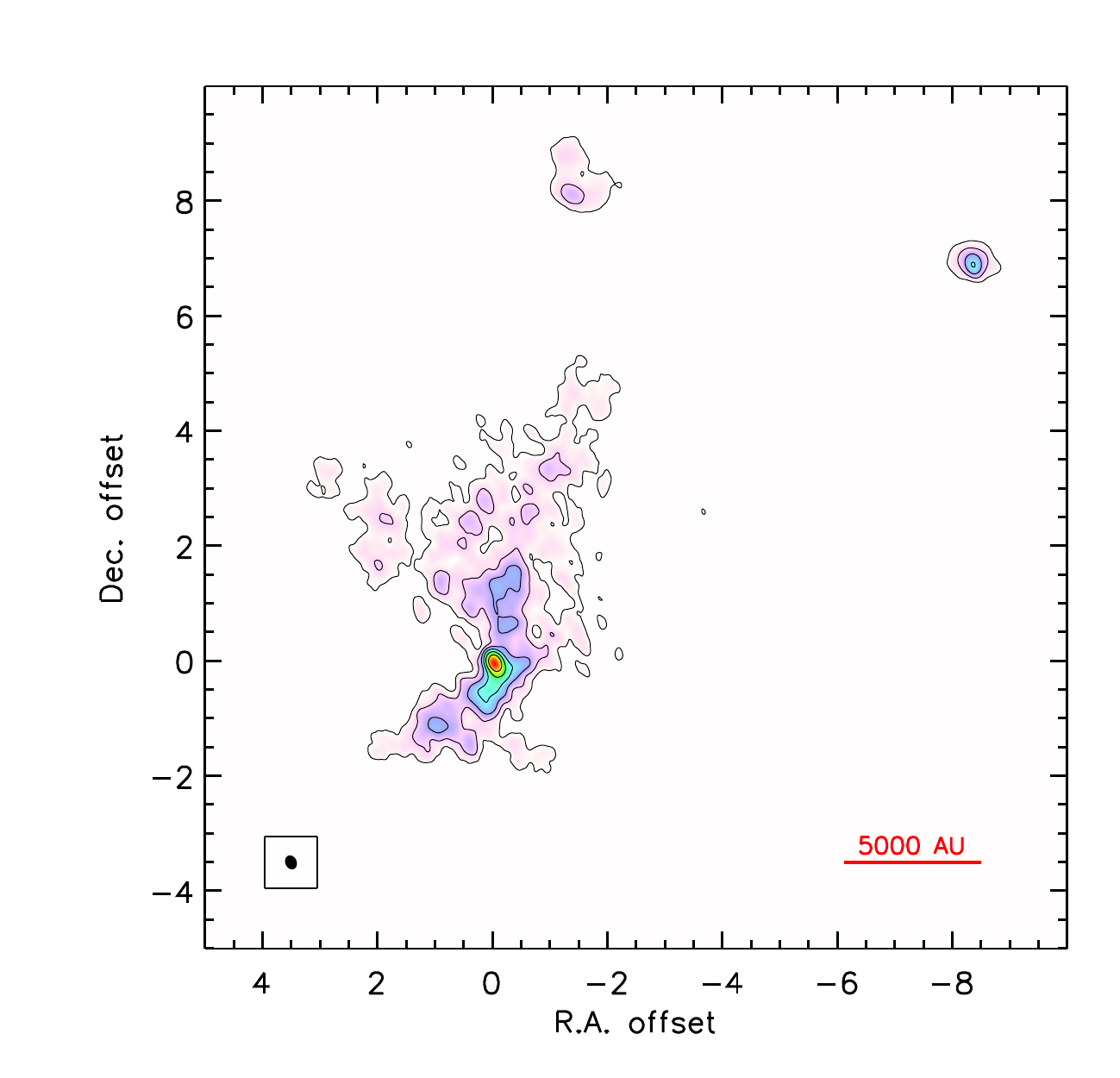}\\
\caption{
1.3 mm continuum map of G339.88-1.26.  The contour levels are
$3\sigma$, $6\sigma$, $12\sigma$, $24\sigma$, ..., with
$1\sigma=0.43~\mJybeam$.  The synthesized beam is $0.29\arcsec\times
0.21\arcsec$ with $\pa=28.2^\circ$.  The R.A. and Dec. offsets are
relative to the continuum peak position
$(\alpha_{2000},\delta_{2000})=(16\h52\m04\s.663$,
$-46^\circ08\arcmin33\arcsec.88)$.}
\label{fig:contmap}
\end{center}
\end{figure}

Figure \ref{fig:contmap} shows the 1.3~mm continuum emission of G339.
This reveals one compact main source at the center with extended
emission, and two separate sources: one in the northwest (about
$11\arcsec$, i.e., $2.3\times10^4~\au$, from the main source); and one
in the north (about $8\arcsec$, i.e., $1.7\times10^4~\au$, from the
main source).  The extended emission associated with the main source
appears to have substructures. It is elongated mostly in the
north-south direction.  Slightly to the south and connected to the
central source, there is an elongated structure in the direction of
northwest-southeast, which may be affected by the outflow (see
\S\ref{sec:outflowtracer}).  The source in the northwest is also
compact. We identify it as a protostar with outflow activities (see
\S\ref{sec:outflowtracer}). The source in the north, however, appears
to be not so compact, and may be part of the extended emission
associated with the main source, especially as the even more extended
emission is resolved out by the interferometric observation.

The total flux of the continuum emission above $3\sigma$ within
$6\arcsec$ from the central source is 0.71~Jy.
To estimate the gas mass from the continuum flux,
we adopt a dust opacity of \citealt[]{Ossenkopf94}
($\kappa_\mathrm{1.3mm}=0.899~\mathrm{cm}^2~\mathrm{g}^{-1}$), and a
gas-to-dust mass ratio of 141 (\citealt[]{Draine11}).
While a temperature of 30 K is typically used for molecular cores,
radiative transfer simulations show that the average temperature of the
envelope within $\sim 10,000~\au$ around a $12-16~M_\odot$ protostar 
(see \S\ref{sec:target}) is $\sim 70~\K$ (\citealt[]{ZT18}). 
The resultant gas mass is $58\:M_\odot$ assuming a temperature of 30~K,
or $23~M_\odot$ assuming 70~K.
This is most likely to be a lower limit for the gas mass due to the resolving out
of more extended emission, which suggests that there is enough
material for future growth of the massive protostar. 
The total flux of the continuum emission within 0.5$\arcsec$ (1,000 au),
i.e., the immediate compact structure around the central protostar,
is 0.16~Jy, which corresponds to a gas mass of $13\:M_\odot$,
assuming a dust temperature of 30 K, 
$5\:M_\odot$ with 70 K, or $1.6~M_\odot$ with 200 K, 
considering even higher temperatures close to the protostar (\citealt[]{ZT18}).
The continuum
fluxes of the sources in the north and northwest are 0.032 and
0.027~Jy, respectively, which correspond to gas masses of 2.2 and
$2.6\:M_\odot$ assuming a dust temperature of 30 K, or 0.9 and
$1\:M_\odot$ assuming a temperature of 70 K.

\subsection{Outflow Tracers}
\label{sec:outflowtracer}

\begin{figure}
\begin{center}
\includegraphics[width=\columnwidth]{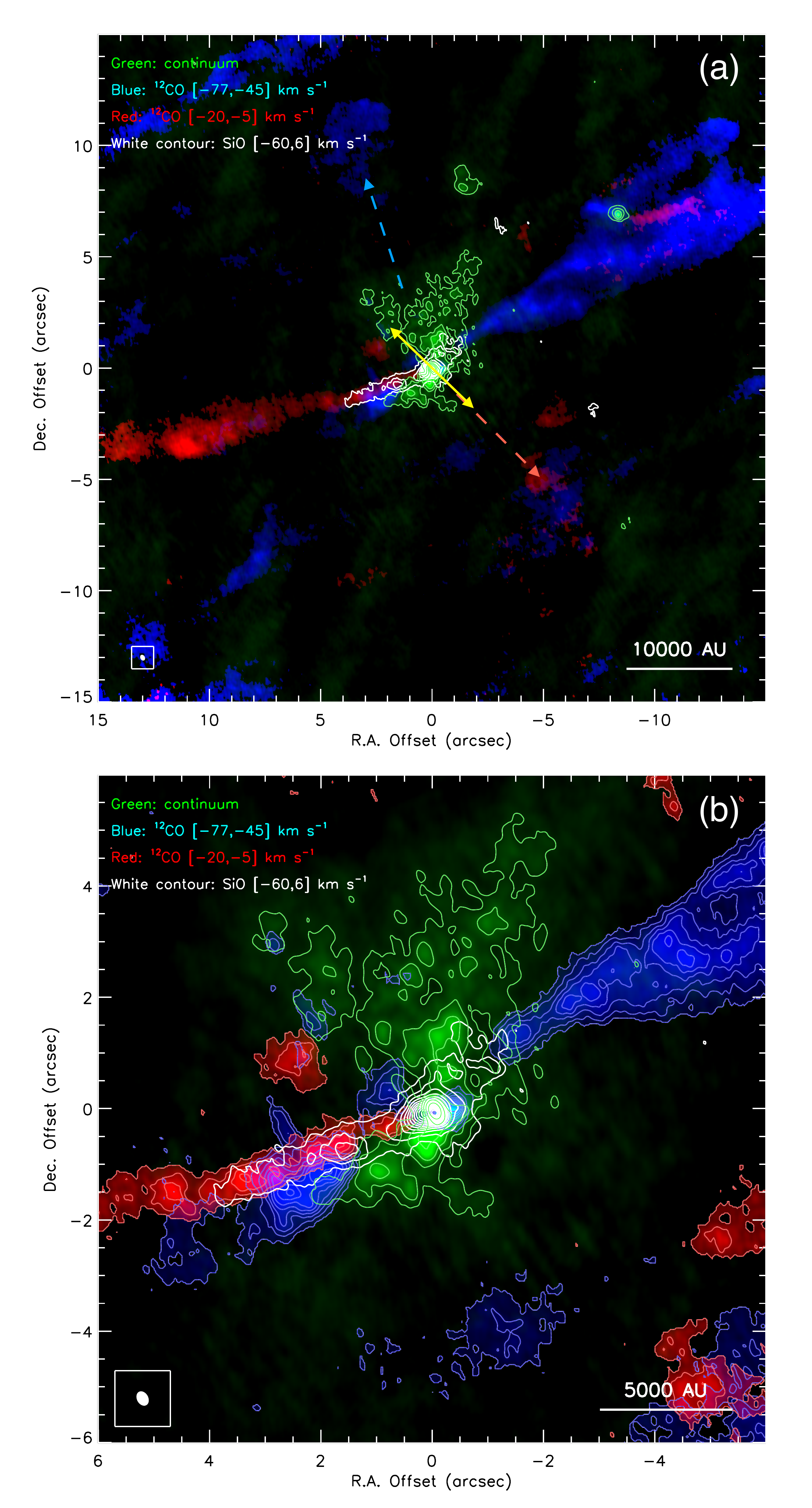}\\
\caption{
{\bf (a):} Integrated maps of the blue-shifted and red-shifted
$^{12}$CO $(2-1)$ emissions in blue and red color scales, overlaid
with the integrated SiO $(5-4)$ emission map (white contours) and 1.3
mm continuum map (green color scale and contours).  The blue-shifted
$^{12}$CO emission is integrated in the range from $\vlsr=-77$ to
$-45~\kms$, while the red-shifted $^{12}$CO emission is integrated in
the range from $\vlsr=-20$ to $-5~\kms$.  The SiO emission is
integrated in the range from $\vlsr=-60$ to $6~\kms$.  The white
contours start at $10\sigma$ and have intervals of $10\sigma$
($1\sigma=20~\mJybeam~\kms$).  The green contours for the continuum
emission are same as those in Figure \ref{fig:contmap}.  The dashed
blue and red arrows mark a possible second outflow from the main
source. The yellow arrows mark the direction of the radio jet discovered in this region.
 {\bf (b):} A zoom-in view of panel (a). The blue and red
contours start at $3\sigma$ and have intervals of $3\sigma$
($1\sigma=24$ and $17~\mJybeam~\kms$ for blue and red contours,
respectively).}
\label{fig:intmap_CO}
\end{center}
\end{figure}

Figure \ref{fig:intmap_CO}(a) shows the large scale $^{12}$CO($2-1$)
emission of the region, revealing a collimated bipolar outflow
associated with the G339 main source.  The red-shifted outflow
emission is detected up to $\vlsr=-5~\kms$ (outflow velocity of
$\vout\equiv\vlsr-\vsys=28~\kms$), while the blue-shifted outflow
emission is detected up to $\vlsr=-77~\kms$ ($\vout=-44~\kms$).  In
the integrated emission map (Figure \ref{fig:intmap_CO}a), the eastern
and western outflows appear to have different opening angles and
directions.  However, as the $^{12}$CO channel maps (Appendix Figure
\ref{fig:chanmap_12CO}) show, in low-velocity channels (e.g.,
$\vlsr=-43$ and $-39~\kms$), the eastern and western outflow lobes
appear to have similar opening angles and are quite well aligned along
a common axis.  Therefore, we estimate that the large-scale CO outflow
has a half opening angle of about $20^\circ$ with a position angle of
about $120^\circ$ east to the north on both sides (indicated by the red dashed lines
in Appendix Figure \ref{fig:chanmap_12CO}).  Furthermore,
although the eastern outflow is dominated by red-shifted emission and
the western outflow is dominated by blue-shifted emission, there are
both blue-shifted and red-shifted emissions on both sides at low
velocities, which are best seen in the channels of $\vlsr=-43$ and
$-23~\kms$ (i.e., $|\vout|\approx 10~\kms$).  Detections of both red
and blue-shifted emissions on both sides suggest a near edge-on view of
this outflow, i.e., the inclination angle between the outflow axis and
the plane of sky is likely to be small (similar to or smaller than the
half opening angle, i.e., $i\lesssim 20^\circ$).

An outflow from this source in roughly the east-west direction was not detected before.
The direction of this CO outflow is actually similar to that of the
elongated MIR emission (\citealt[]{Debuizer02}; see \S\ref{sec:multiwave}). 
The direction of
this outflow is also similar to that of the CH$_3$OH maser
distribution, which suggests that these masers may trace the shocks
produced in outflow activities (see discussions in \S\ref{sec:chemistry}). 
There is tentative indication of a second molecular
outflow from the main source with position angles of $18^\circ$
(blue-shifted) and $-135^\circ$ (red-shifted) (labeled with dashed
arrows in Figure \ref{fig:intmap_CO}a).  In the channel maps at low
velocities (e.g., $\vlsr=-43$ and $-39~\kms$), there are also
small-scale emissions close to the main source with a position angle
of about $45^\circ$.
These emissions may belong to the molecular outflow
associated with the radio jet previously observed
(\citealt[]{Ellingsen96,Purser16}) which has a position angle of
$46^\circ$ (indicated by yellow arrows in Figure \ref{fig:intmap_CO}a).
If this is the case, it indicates that
there is an embedded proto-binary system with individual outflows
almost perpendicular to each other.  However, at current resolution
($\sim 0.3\arcsec$), this binary is still unresolved.  For the
continuum source located in the north-west of the main source, a small
collimated bipolar outflow is detected.  There is no clear outflow
emission associated with the continuum source north of the main
source.

Figure \ref{fig:intmap_CO}(b) shows a zoom-in view of the CO outflow
emission.  The SiO $(5-4)$ emission shows a highly collimated jet
extending from the continuum peak to about $5\arcsec$ away.  The jet
structure in the SiO only shows emission in the red-shifted velocities
ranging from $\vlsr=-33$ to $-7~\kms$ (see Appendix Figure
\ref{fig:chanmap_SiO}).  The position angle of the SiO jet is about
$110^\circ$, slightly different from the large-scale 
$^{12}$CO outflow with a position angle of $120^\circ$.
However, the red-shifted $^{12}$CO emission on the same scale of
the SiO jet coincides very well with the SiO emission.
This may be caused by the jet precession. 
The width of the SiO jet is about
$0.6\arcsec$, and there is no apparent change in the jet width with
distance to the source.

\subsection{Envelope/Disk Tracers}
\label{sec:envelopetracer}

\begin{figure*}
\begin{center}
\includegraphics[width=0.85\textwidth]{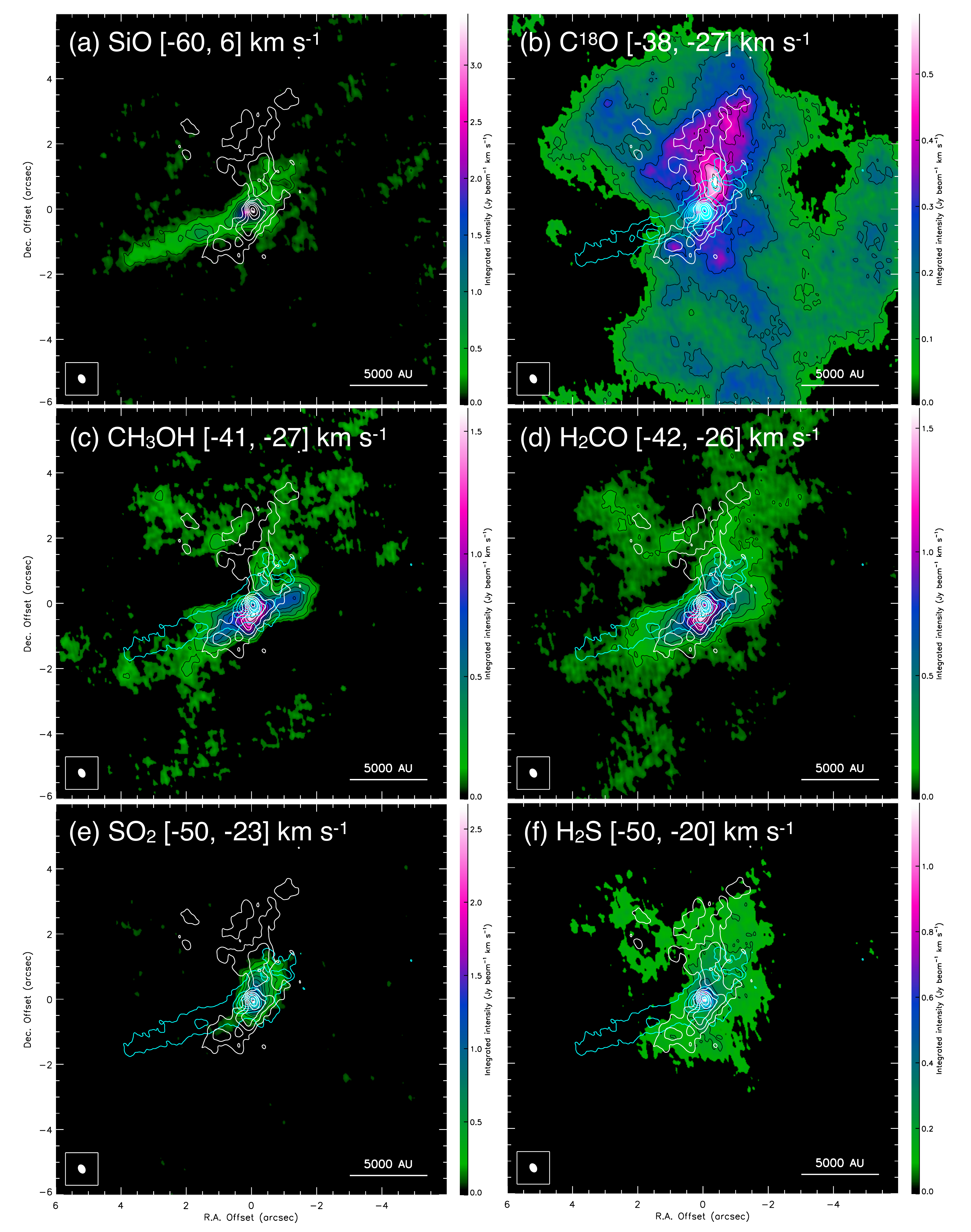}\\
\caption{
Integrated emission maps of SiO (panel a), C$^{18}$O (panel b),
CH$_3$OH (panel c), H$_2$CO (panel d), SO$_2$ (panel e), and H$_2$S
(panel f) shown in color scales and black contours. The velocity
ranges for the integrated maps are labelled in each panel. The black
contours start from $10\sigma$ and have intervals of $10\sigma$
($1\sigma=20$, 9.4, 14, 12, 19, and $17~\mJybeam~\kms$ in panels
a$-$f, respectively).  The continuum emission is shown in the white
contours with contour levels of $5\sigma$, $10\sigma$, $20\sigma$,
$40\sigma$, ..., with $1\sigma=0.43~\mJybeam$. The cyan contours in
panels (b)$-$(f) show the integrated SiO emission (same as panel a)
for reference.}
\label{fig:intmap}
\end{center}
\end{figure*}

\begin{figure}
\begin{center}
\includegraphics[width=\columnwidth]{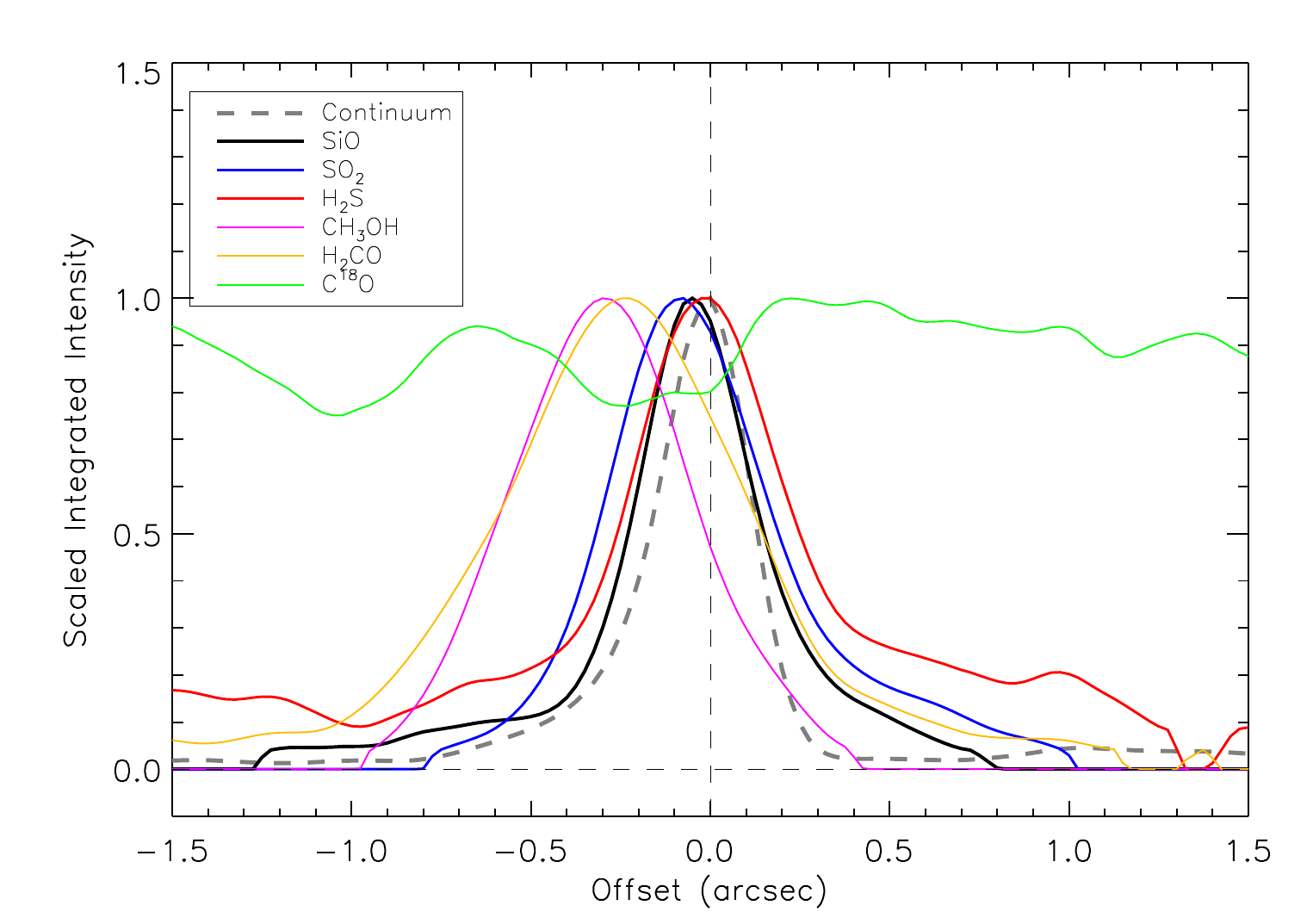}\\
\caption{
The intensity profiles of the integrated SiO, SO$_2$, H$_2$S,
CH$_3$OH, H$_2$CO, and C$^{18}$O emissions (solid lines), and the
continuum emission (dashed lines).  The profiles are extracted along a
cut perpendicular to the SiO jet and through the continuum peak
($\pa=20^\circ$). The position offsets are relative to the continuum
peak position. The intensity profiles are normalized by their maximum
values.}
\label{fig:fluxprofile}
\end{center}
\end{figure}

\begin{figure*}
\begin{center}
\includegraphics[width=0.85\textwidth]{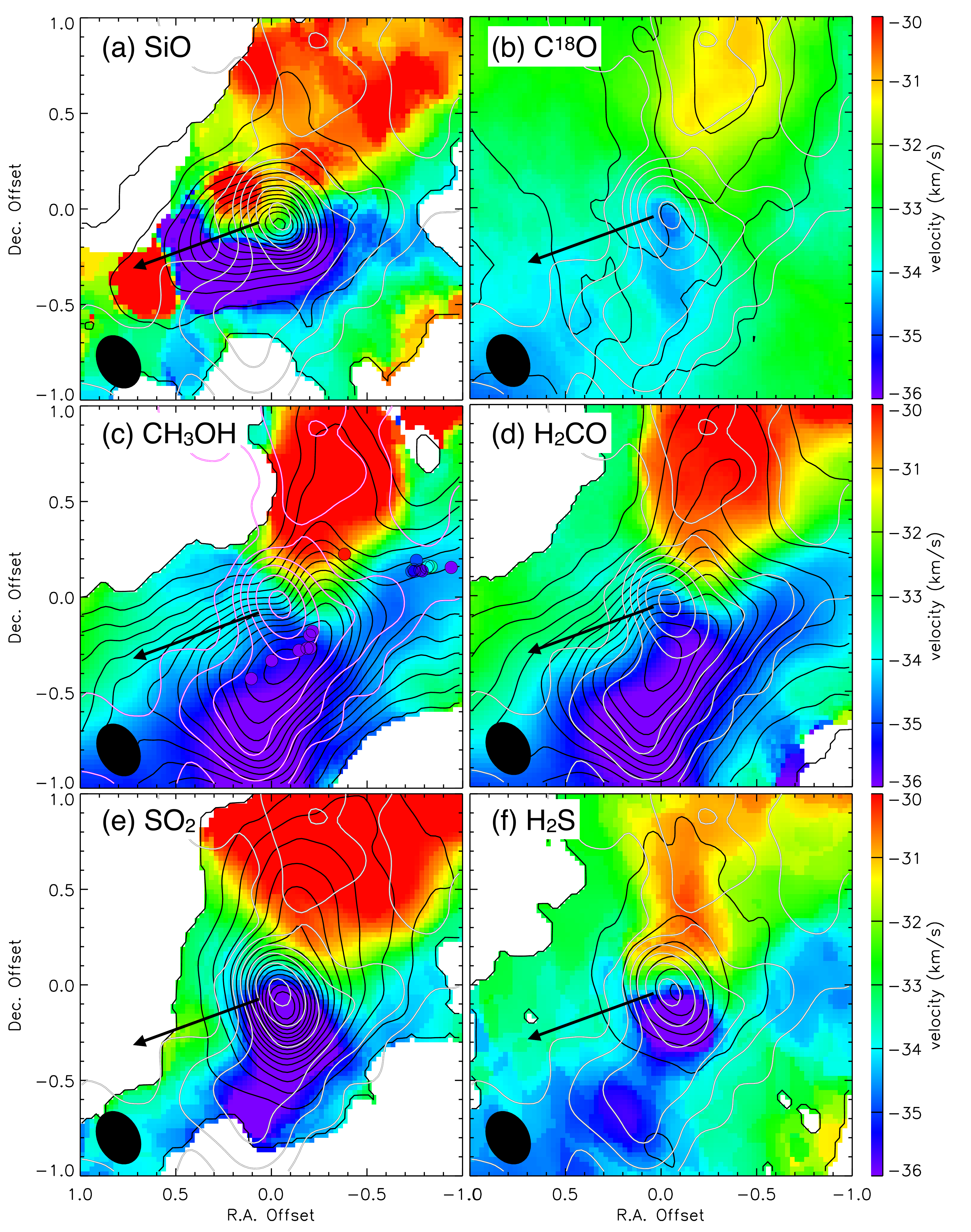}\\
\caption{
Moment 0 maps (black contours) and moment 1 maps (color scale) of the
SiO (panel a), C$^{18}$O (panel b), CH$_3$OH (panel c), H$_2$CO (panel
d), SO$_2$ (panel e), and H$_2$S (panel f) emissions.  The black
contours start from $5\sigma$ and have intervals of $10\sigma$
($1\sigma=20$, 9.4, 14, 12, 19, and $17~\mJybeam~\kms$ in panels
a$-$f, respectively). The continuum emission is shown in the white
contours with contour levels of $5\sigma$, $10\sigma$, $20\sigma$,
$40\sigma$, ..., with $1\sigma=0.43~\mJybeam$.
The black arrow in each panel indicates the direction of the SiO jet.
The circles in panel c show the positions and velocities (in color scale) 
of the CH$_3$OH masers from \citet[]{Dodson08}.}
\label{fig:momentmap}
\end{center}
\end{figure*}

\begin{figure*}
\begin{center}
\includegraphics[width=\textwidth]{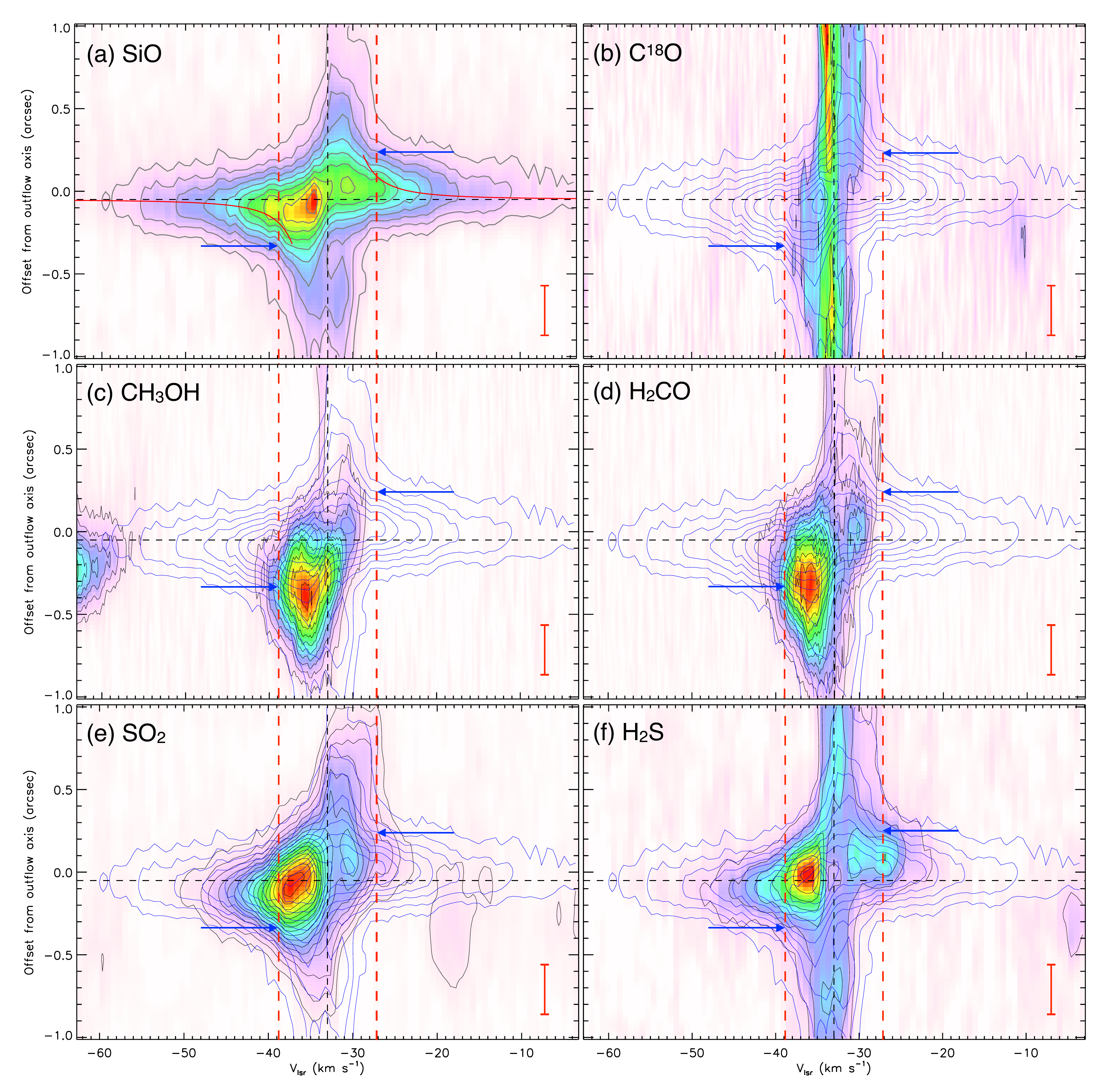}\\
\caption{
Position-velocity diagrams of the SiO (panel a), C$^{18}$O (panel b),
CH$_3$OH (panel c), H$_2$CO (panel d), SO$_2$ (panel e), and H$_2$S
(panel f) emissions along cuts perpendicular to the SiO jet and
through the continuum peak ($\pa=20^\circ$) shown in color scales and
grey contours.  The width of the cuts is $0.3\arcsec$.  The grey
contours start from $5\sigma$ and have intervals of $5\sigma$
($1\sigma=2.3$, 5.6, 4.2, 4.0, 2.4, and $2.4~\mJybeam$ in panels
(a)$-$(f), respectively).  In panels (b)$-$(f), the blue contours show
the position-velocity diagram of the SiO emission (same as panel a)
for reference.  The position offsets are relative to the continuum
peak position. The horizontal dashed line marks the position of the
peak position of the integrated SiO emission peak.  The black vertical
line is the systemic velocity of the source ($\vsys=33~\kms$).  The
red vertical lines mark the rotation velocity at the centrifugal
barrier and the blue arrows mark its position (see text).  The red
curves in panel (a) are the Keplerian rotation curve with the central
mass estimated from the velocity and radius of the centrifugal
barrier.  The red bar at the lower-right corner of each panel
indicates the resolution beam size.}
\label{fig:pvdiagram}
\end{center}
\end{figure*}

\begin{figure}
\begin{center}
\includegraphics[width=\columnwidth]{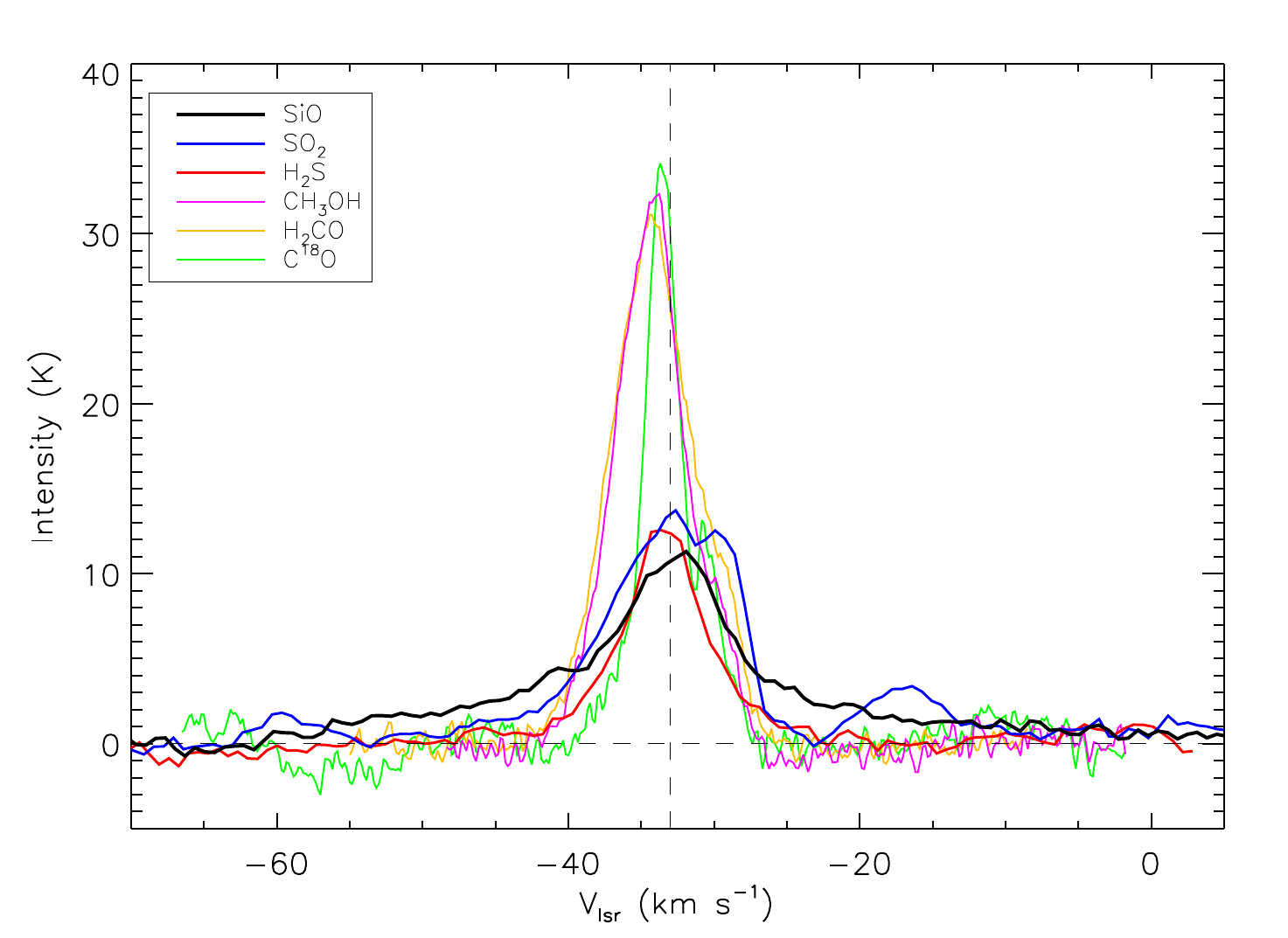}\\
\caption{
The spectra of SiO ($5-4$), SO$_2$ ($22_{2,20}-22_{1,21}$),
H$_2$S ($2_{2,0} - 2_{1,1}$), CH$_3$OH ($4_{2,2}-3_{1,2}$; E),
H$_2$CO ($3_{2,1} - 2_{2,0}$), and C$^{18}$O ($2-1$) lines within
$1\arcsec$ from the position of continuum peak. The vertical line is
the systemic velocity of the source ($\vsys=33~\kms$).
The emission at $\vlsr\sim -16~\kms$ in the SO$_2$ spectrum
is due to the CH$_3$CHO ($11_{1,10}-10_{1,9}$; E) emission.}
\label{fig:spectrum}
\end{center}
\end{figure}

Figure \ref{fig:intmap} shows the integrated emission maps of
SiO($5-4$), C$^{18}$O($2-1$), CH$_3$OH($4_{2,2}-3_{1,2}$; E),
H$_2$CO($3_{2,1} - 2_{2,0}$), SO$_2$($22_{2,20}-22_{1,21}$), and
H$_2$S($2_{2,0} - 2_{1,1}$) lines.  The SiO emission not only traces
the jet but is also strongly peaked at the continuum source.  The SiO
emission peak coincides very well with the continuum peak, with an
elongation in the direction of the jet.  The SiO emission at the
continuum peak is detected within a wide velocity range from
$\vlsr=-60$ to $6~\kms$ (i.e., up to about $30-40~\kms$ relative to the
systemic velocity).

The spatial distributions of the other molecular line emissions can be
categorized into three types.  First, the SO$_2$ and H$_2$S emissions
are strongly peaked at the positions of the continuum source and the
SiO emission peak. The SO$_2$ emission is only seen within $\sim
1\arcsec$ (2100~au) from the central source, while the H$_2$S emission
also traces the extended structure up to $\sim 4\arcsec$ (8400~au)
from the central source.  The emissions of these two molecules at the
continuum peak are detected in a velocity range smaller than the SiO
emission (up to about $20~\kms$ relative to the systemic velocity; see
Appendix Figures \ref{fig:chanmap_SO2} and \ref{fig:chanmap_H2S}).
Second, the CH$_3$OH and H$_2$CO emissions have very similar
behaviors.  They are both detected in a narrower velocity range than
the SO$_2$ and H$_2$S emissions (up to about $10~\kms$ relative to the systemic velocity;
see Appendix Figures \ref{fig:chanmap_CH3OH} and
\ref{fig:chanmap_H2CO}).  Their emission peaks are offset from the
continuum peak and the SiO emission peak.  On the larger scale, they
generally follow the morphology of the extended continuum emission.
In addition to that, they also show strong emissions associated with
the outflow cavity, which is most clearly seen in the CH$_3$OH
emission, showing a cone-like structure surrounding the SiO jet.
Third, the C$^{18}$O emission is widely spread. Its emission peak is
$\sim 1\arcsec$ to the north of the continuum peak, corresponding to a
separate peak in the extended continuum emission. The large scale
C$^{18}$O emission follows the morphology of the continuum emission,
especially to the north. However, it is much more widely distributed
than the continuum emission.  The C$^{18}$O emission is only detected
within a narrow velocity range up to about $5~\kms$ relative to the
systemic velocity (see Appendix Figures \ref{fig:chanmap_C18O}).

Figure \ref{fig:fluxprofile} shows the intensity profiles of the
integrated emissions of these lines, as well as the continuum
emission, along a cut passing through the continuum peak and
perpendicular to the SiO jet (i.e., $\pa=20^\circ$).  This further
confirms the features discussed above.  We see that SiO, SO$_2$ and
H$_2$S emissions are highly peaked at positions close to the central
source, while the peaks of the CH$_3$OH and H$_2$CO emissions are more offset from
the central source ($\sim0.3\arcsec$). On the other hand, the
C$^{18}$O emission is quite widespread across the region.

\section{Rotating Envelope-Disk}
\label{sec:envelope}

\subsection{Velocity Structure}
\label{sec:rotation}

Figure \ref{fig:momentmap} shows the moment 1 maps of SiO($5-4$),
C$^{18}$O($2-1$), CH$_3$OH($4_{2,2}-3_{1,2}$; E), H$_2$CO($3_{2,1} -
2_{2,0}$), SO$_2$($22_{2,20}-22_{1,21}$), and H$_2$S($2_{2,0} -
2_{1,1}$) emissions in a region within $1\arcsec$ (2100~au) from the central
source.
Velocity gradients are seen in all of these molecular line
emissions across the central source approximately in north-south direction. 
The magnitude of the velocity gradients are
different in these molecular emissions. 
The SiO and SO$_2$ emissions appear to
have the strongest velocity gradients confined in a region close to the
central source ($< 0.5\arcsec$, 1000 au).
The H$_2$S emission has a strong velocity gradient on similar scales,
but also a smaller gradient on larger scales.  On the small scale, the
velocity gradients in CH$_3$OH and H$_2$CO are smaller than those in SiO
or SO$_2$, but velocity gradients can also be seen on larger scales.
The velocity gradient in the C$^{18}$O emission is the weakest among
these lines. 
The moment 1 maps of these molecular lines for the whole region
are shown in Appendix Figure \ref{fig:momentmap_large}. 
Besides the varying velocity gradient levels, the
detailed directions of the velocity gradients are also different in
these molecular lines.  The velocity gradient in the SiO emission,
as well as those in the SO$_2$ and H$_2$S emissions, 
is mostly perpendicular to the SiO jet axis (i.e. $\pa=20^\circ$,
black arrow in Figure \ref{fig:momentmap}), 
which is consistent with rotation. 
The velocity gradients in the CH$_3$OH and H$_2$CO
emissions, however, also have components along the direction of the outflow
axis, which may indicate infalling motion in the envelope,
and/or outflow motion, in addition to rotation. 

To better show the velocity structures, in Figure \ref{fig:pvdiagram}
we show the position-velocity (PV) diagrams of these molecular line
emissions, along a cut passing through the continuum peak and
perpendicular to the SiO jet (i.e., $\pa=20^\circ$). 
Signatures consistent with rotation across the jet axis
are seen in all these molecular lines,
with the southern side dominated by the blue-shifted emissions,
and the northern side dominated by the red-shifted emissions.
Note that, from the PV diagram of the SiO emission (panel a),
the offsets are symmetric to a position slightly south of the continuum peak position
by $0.05\arcsec$ (marked by the horizontal lines in Figure
\ref{fig:pvdiagram}), which is much smaller than the resolution beam
size ($0.3\arcsec$).
However, while the SiO emission probes material with high rotation velocities
up to $\sim 30~\kms$ close to the position of the central source,
the highest velocities detected in the CH$_3$OH and H$_2$CO emissions 
are $<10~\kms$ and offset from the central source in space.
The kinematics of the SO$_2$ and H$_2$S emissions shown in the PV
diagrams appear to be in between those of SiO and those of CH$_3$OH
and H$_2$CO. In velocity space, they show velocities higher than those
of CH$_3$OH and H$_2$CO, but not as high as those of the SiO emission.
Spatially, the main parts of the SO$_2$ and H$_2$S
emissions are more confined than the CH$_3$OH and H$_2$CO emissions.
Note that the H$_2$S emission also contains an extended component with very
low velocities.  
The C$^{18}$O emission only shows a very weak
velocity gradient.  The positional offsets of the most
blue and red-shifted C$^{18}$O emissions are also consistent with those seen
in the CH$_3$OH and H$_2$CO PV diagrams.

Figure \ref{fig:spectrum} shows the spectra of these
lines within a radius of $1\arcsec$ from the continuum peak position.
The SiO line clearly shows strong
high-velocity wings, which are not seen in other molecular lines.  The
SO$_2$ and H$_2$S lines are narrower than
that of SiO, but wider than the other lines.
Note that the emission at $\vlsr\sim -16~\kms$ in the SO$_2$ spectrum
is due to the CH$_3$CHO ($11_{1,10}-10_{1,9}$; E) emission.
The CH$_3$OH and H$_2$CO lines, despite being very bright, 
are even narrower.  The C$^{18}$O line is the narrowest among these lines.
The different velocity ranges of these molecular lines are 
from genuinely different velocity components, as we discussed above,
rather than the result of different line intensities.

The different behaviors of these molecular lines in their velocity and spatial 
distributions suggest that they trace different structures around the protostar.
In the CH$_3$OH and H$_2$CO emissions, the highest velocities
are detected offset from the position of the central source.
In addition, on the northern side, while most of the emission is red-shifted, there is
significant blue-shifted emission as well. The opposite is true for
the southern side. 
Meanwhile, velocity gradients are also seen in the PV diagram perpendicular
to the disk direction (see Figure \ref{fig:model}).
Such behaviors are consistent with infalling-rotating motion in the
envelope (e.g., \citealt[]{Sakai14b,Oya15,Oya16,Oya17}), rather than
pure rotation in the disk (see also Appendix \ref{sec:appC}).
The locations of the most blue and red-shifted emissions 
then correspond to the innermost radius of such an envelope
(indicated by the blue arrows in Figure \ref{fig:pvdiagram}), where all
the kinetic energy is converted to rotation (i.e., the centrifugal
barrier; \citealt[]{Sakai14}).  The rotation reaches its maximum
velocity at this point without losing angular momentum. Higher
rotational velocity is not seen in these lines since they do
not trace the disk inside of the envelope.
The SiO, SO$_2$, and H$_2$S  emissions, on the other hand, 
show higher velocities close to the central source, indicating that they
are tracing the disk.
However, the SO$_2$ and H$_2$S emissions have their highest velocities 
lower than those of the SiO emission, suggesting that they may only
trace the outer part of the disk.
Note that the low-velocity components of the SiO, SO$_2$, and H$_2$S emissions
in the PV diagrams are consistent with the CH$_3$OH and H$_2$CO emissions. 
This suggests that they do not only trace the disk, but also some parts of the envelope. 
The C$^{18}$O line has much lower critical density than the other lines (Table \ref{tab:lines}),
so its emission is dominated by the outer low-density material and does not show
high velocities.

To summarize, the spatial distributions and kinematics of these
molecular emissions are consistent with a scenario in which there is a
transition from an infalling-rotating envelope to a disk, and
different molecular emissions trace different components (see also Appendix \ref{sec:appC}).
The SiO emission traces the disk and the inner envelope,
the CH$_3$OH and H$_2$CO emissions trace the envelope, and the
SO$_2$ and H$_2$S emissions trace the outer part of the
disk, as well as the inner envelope. 
Based on such a scenario, we can approximately estimate the
velocity and radius of the centrifugal barrier (marked by the red
dashed lines and blue arrows in Figure \ref{fig:pvdiagram}) to be
$\vcb=6\pm1~\kms/\cos i$ and $\rcb/d=0.25\arcsec\pm0.05\arcsec$, i.e.,
$\rcb=530\pm110~\au$, which lead to a central mass estimate of
$m_{*d}=\rcb\vcb^2/(2G)=11^{+6}_{-5}~M_\odot/\cos^2 i$
(\citealt[]{Sakai14}), where $i$ is the inclination angle defined with
$i=0^\circ$ for an edge-on disk and $i=90^\circ$ for a face-on disk.
Note that the estimated dynamical mass $m_{*d}$ includes both the protostellar mass
and disk mass.
The ratio between the disk mass and the protostellar mass $f_d$ is uncertain.
Theoretical modeling suggests that the disk mass can be a significant
fraction of the protostellar mass ($f_d\sim 1/3$; e.g. \citealt[]{Kratter08}),
while some observations provide estimates of $f_d=1/30-1/15$ (e.g., \citealt[]{Ilee16}).
The estimated dynamical mass is roughly consistent with the mass estimate of
$12-16\:M_\odot$ based on SED fitting (\citealt[]{Liu19}).  With
such an estimated central mass of $11\:M_\odot$,
the red curves in Figure \ref{fig:pvdiagram}(a) show the rotation
curve of a Keplerian disk inside of the centrifugal barrier, assuming a negligible disk mass.
Note that the rotation velocity of the Keplerian disk at the centrifugal barrier
is a factor of $\sqrt{2}$ lower than the rotation velocity of the infalling-rotating
envelope at the centrifugal barrier (see Eq. \ref{eq:vphi_disk}).
The high velocity components of SiO emission are consistent with such a
rotation curve, suggesting a rotationally supported disk inside the centrifugal
barrier.  However, it is difficult to obtain the detailed rotation profile in the disk with the current
angular resolution of the data.

\subsection{Kinematic Model of the Rotating Envelope-Disk}
\label{sec:model}

\begin{figure*}
\begin{center}
\includegraphics[width=0.7\textwidth]{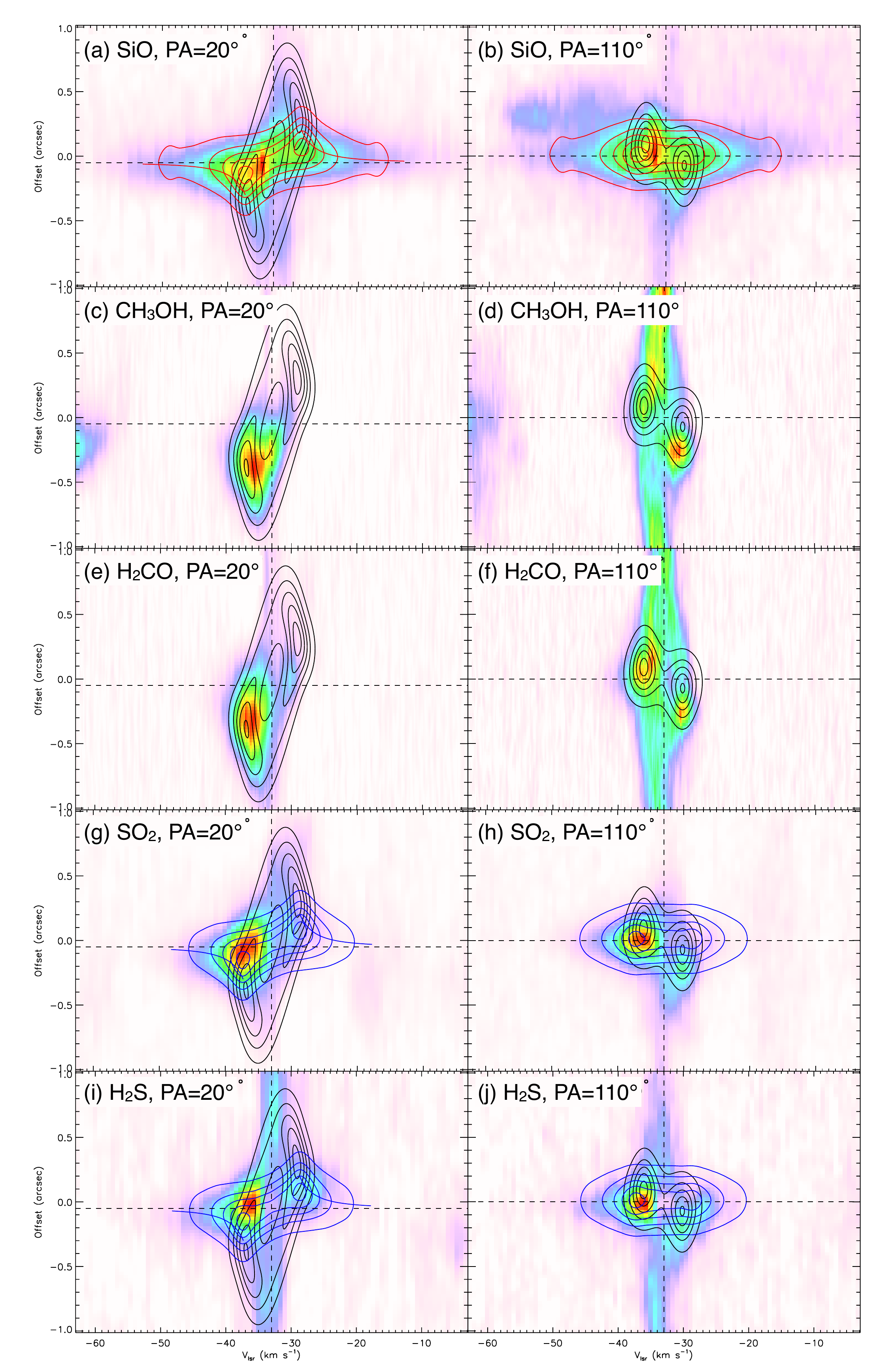}\\
\caption{
Models of infalling-rotating envelope and Keplerian disk compared with
the observations. The left column shows the position-velocity diagrams
of the SiO ($5-4$), CH$_3$OH ($4_{2,2}-3_{1,2}$; E), H$_2$CO ($3_{2,1} -
2_{2,0}$), SO$_2$ ($22_{2,20}-22_{1,21}$), and H$_2$S ($2_{2,0} -
2_{1,1}$) emission along a cut perpendicular to the outflow axis (same
as Figure \ref{fig:pvdiagram}). The positive offsets are to the north
of the source. The right column shows the position-velocity diagrams
of same lines along a cut along the outflow axis.  The positive
offsets are on the east side of the source.  The observed data are
shown in the color scale. The black contours show the model of an
infalling-rotating envelope outward of the centrifugal barrier.  The
red contours show the model of a Keplerian disk inward of the
centrifugal barrier.  The blue contours show the model of the same
Keplerian disk but with the inner region truncated at a particular
radius. The model contours are at levels of 0.1, 0.3, 0.5, 0.7, and 0.9 of the peak intensities.
The red curves in panel (a) and the blue curves in panels (g) and (i) 
are the Keplerian rotation curve for the disk models.}
\label{fig:model}
\end{center}
\end{figure*}

In order to better illustrate the changes of kinematics and types of molecular line emissions
in the transition from envelope to disk, we construct
simple models to compare with the observed PV diagrams. In order to
separate the kinematic feature of an infalling-rotating envelope and a
disk, we construct models for these two components separately. In the model, the
envelope starts from the centrifugal barrier at radius of $\rcb$ as
its inner boundary and extends to an outer boundary $\rout$, with the
rotation velocity $v_\varphi$ and infall velocity $v_r$ described as
\begin{eqnarray}
v_\varphi & = & \vcb \frac{\rcb}{r},\\
v_r & = & -\vcb\frac{\sqrt{\rcb(r-\rcb)}}{r}.
\end{eqnarray}
Such motion conserves both angular momentum and mechanical energy.
$\rcb$ is the innermost radius that such infalling gas can reach with
the angular momentum conserved. At $r=\rcb$, $v_r=0$ and
$v_\varphi=\vcb$.  Since we only focus on the kinematic features in
this paper, we adopt simple geometry and density structures in the
model. We assume the envelope has a height of $h(r)=0.2r$ on each side
 of the mid-plane, and the density distribution follows $\rho(r)\propto
r^{-1.5}$.  For simplicity, we assume the emissions are optically thin
and the excitation conditions are universal across the
region. 
We also fix $\rout/d=0.8\arcsec$ ($\rout=1.7\times10^3~\au$) based on the observed PV
diagrams. Therefore, we have in total three free parameters for the
model of infalling-rotating envelope: the radius of the centrifugal
barrier, $\rcb$; the rotation velocity at the centrifugal barrier,
$\vcb$; and the inclination angle, $i$.  In such a model, the central
mass is $m_{*d}=\rcb\vcb^2/(2G)$.  

For the disk inside of the envelope,
we assume that it is Keplerian and it has the centrifugal barrier as
its outer boundary.  The rotation velocity $v_\varphi$ and infall
velocity $v_r$ in the disk are
\begin{eqnarray}
v_\varphi & = & \sqrt{\frac{Gm_*}{r}}= \sqrt{\frac{\rcb\vcb^2}{2r}},\label{eq:vphi_disk}\\
v_r & = & 0.
\end{eqnarray}
Here we assumed $m_*=m_{*d}$.
We assume the disk also has a height of $h(r)=0.2r$ on each side of
the mid-plane, and a density profile of $\rho(r)\propto r^{-2.5}$.
Similarly, we also assume the emissions are optically thin and the
excitation conditions are universal across the region.

To obtain the best-fit model, we compare the model PV diagrams of the
infalling-rotating envelope with the PV diagrams of the CH$_3$OH,
H$_2$CO, and the outer part of the SiO emissions, and compare the
model PV diagram of the Keplerian disk with the PV diagram of the
inner part of the SiO emission. We explore the inclination angle $i$
with values ranging from $0^\circ$ to $40^\circ$ with an interval of
$10^\circ$, the angular radius of centrifugal barrier $\rcb/d$ in a range of
$0.1\arcsec - 0.4\arcsec$ with an interval of $0.05\arcsec$, and the
projected centrifugal barrier velocity $\vcb\cos i$ in a range of
$3-8~\kms$ with an interval of $1\:\kms$, and approximately determine
the best-fit model by eye (given the approximate nature of the
modeling).

The best-fit model has an inclination angle of $i=10^\circ$ between
the line of sight and the disk mid-plane, a radius of the centrifugal
barrier of $\rcb/d=0.25\arcsec$, i.e., $\rcb=530~\au$, and a velocity at the
centrifugal barrier of $\vcb\cos i=6~\kms$.  The central mass is
therefore derived to be about $11~M_\odot$.  The best model is shown
in Figure \ref{fig:model}.  It shows that the CH$_3$OH and H$_2$CO
emissions indeed can be explained as an infalling-rotating envelope,
without any high-velocity component associated with the disk rotation.
There are extended emissions in the direction of the outflow axis
(panels d and f in Figure \ref{fig:model}), which may be affected by
the outflow cavity structure, as mentioned in
\S\ref{sec:envelopetracer}. The high-velocity SiO emission is
consistent with a Keplerian disk, while the low-velocity SiO emission
in the outer region is consistent with the infalling-rotating
envelope.  Especially, the velocity gradient seen along the outflow
axis direction across the central source (panel b) cannot be explained
by pure rotation in the disk.  

The SO$_2$ and H$_2$S emissions show
some high-velocity components associated with the disk, but the
highest velocity is not as high as the SiO emission, suggesting that
they only trace the outer part of the disk.
Therefore, we construct another disk model with its inner part
truncated at a certain radius $r_\mathrm{in}$ for modeling
the SO$_2$ and H$_2$S emissions. We explore the inner
radius of the disk $r_\mathrm{in}/d$ in the range of
$0.01\arcsec-0.15\arcsec$ with an interval of $0.01\arcsec$.
The other parameters of this model are same as the best model obtained above.
By comparing the PV diagrams of the model and observation, we estimate
$r_\mathrm{in}/d=0.02\arcsec$, i.e. $r_\mathrm{in}=$40 au.  Meanwhile, there are
some extended low-velocity SO$_2$ and H$_2$S emissions which may be
associated with the infalling-rotating envelope.

To summarize, the model supports our hypothesis that the different
molecular emissions trace different parts of the transition from an
infalling-rotating envelope to a Keplerian disk.  The CH$_3$OH and
H$_2$CO emissions trace the infalling-rotating envelope outside of the
centrifugal barrier at a radius of 530 au. The SiO emission also
traces the Keplerian disk inside of the centrifugal barrier, in
addition to the envelope. The SO$_2$ and H$_2$S emissions trace
the centrifugal barrier and the disk inward but
outside of a radius of about 40 au. 
In Appendix \ref{sec:appC}, we discuss the possibility that all these emissions
trace different parts of a single rotationally supported disk without infall motion.

\subsection{The Model Caveats}
\label{sec:caveats}

We note that the model presented above is only a simple example that
is designed to be illustrative.  It is focused on explaining the
kinematic features seen in different molecular lines. In reality, the
geometry of the structures, including the density and temperature
distributions, are likely to be more complicated than our model
assumptions. 
The observed change of the types of molecular line emissions in the transition
from the envelope to the disk is not only a result of the change of chemical composition,
but is also affected by the change of excitation conditions (see \S\ref{sec:chemistry}).
Also, unlike the model in which the molecules are
uniformly distributed in the two components (envelope and disk), these
molecules are likely to have more complicated abundance distributions,
including vertical differentiations.

In the model, we also ignore the motions of the infalling material in
the $z$ direction (i.e., perpendicular to the envelope/disk
mid-plane).  In ballistic models of infall (e.g.,
\citealt[]{Ulrich76,Cassen81,Stahler94}), material streams land on the
mid-plane at their centrifugal radii.  These streams will collide and
form a disk structure, with its outer boundary set by the centrifugal
radius of the material infalling along the mid-plane.  Material in
this disk structure will continue to spiral inward with both rotation
and infall until reaching the centrifugal barrier, inside of which a
rotationally supported disk forms (\citealt[]{Stahler94}).  Therefore,
a pseudo disk dominated by infalling-rotating motion can form between
the envelope and the rotationally supported disk (e.g.,
\citealt[]{Lee14}).  In this paper, however, by only considering the
motion along the envelope/disk mid-plane, we do not distinguish an
infalling-rotating pseudo disk and an infalling-rotating envelope.
Furthermore, the disk formation process can be strongly affected by
the magnetic field, by removing angular momentum via the magnetic
braking effect (e.g., \citealt[]{Li14,Zhao16}). Such effects are also
beyond the scope of our simple model.

In the model, the velocity gradient seen in the PV diagram along the
outflow axis (i.e., perpendicular to the disk direction) is caused by
the infalling motion of the envelope.  It is slightly blue-shifted on
the eastern side and red-shifted on the western side.  In such a case,
the back side of the envelope is on the east side and the front side
of the envelope is on the west side.  That is to say, if an outflow is
perpendicular to this envelope, it will be blue-shifted on the east
and red-shifted on the west, which is not consistent with the observed
jet.  However, this misalignment is actually small considering the
near edge-on view of both the envelope and the outflow (see
\S\ref{sec:outflowtracer}).  It is possible that there is a small
misalignment between the envelope and the disk/jet.  Changes of
angular momentum direction during the accretion process are possible
considering that the core is highly sub-structured and expected in the
collapse of a turbulent core.  It is also possible that the direction
of the jet may have changed modestly over the time, as discussed in
\S\ref{sec:outflowtracer}.  In fact, there are some distinct
blue-shifted SiO outflow emissions to the east of the central source.
which can be seen in Figure \ref{fig:model}(b).  If that is the most
recently launched jet, it is consistent with the inclination of the
envelope.

\section{Outflow}
\label{sec:outflow}

\subsection{The SiO Jet}
\label{sec:launch}

\begin{figure*}
\begin{center}
\includegraphics[width=0.7\textwidth]{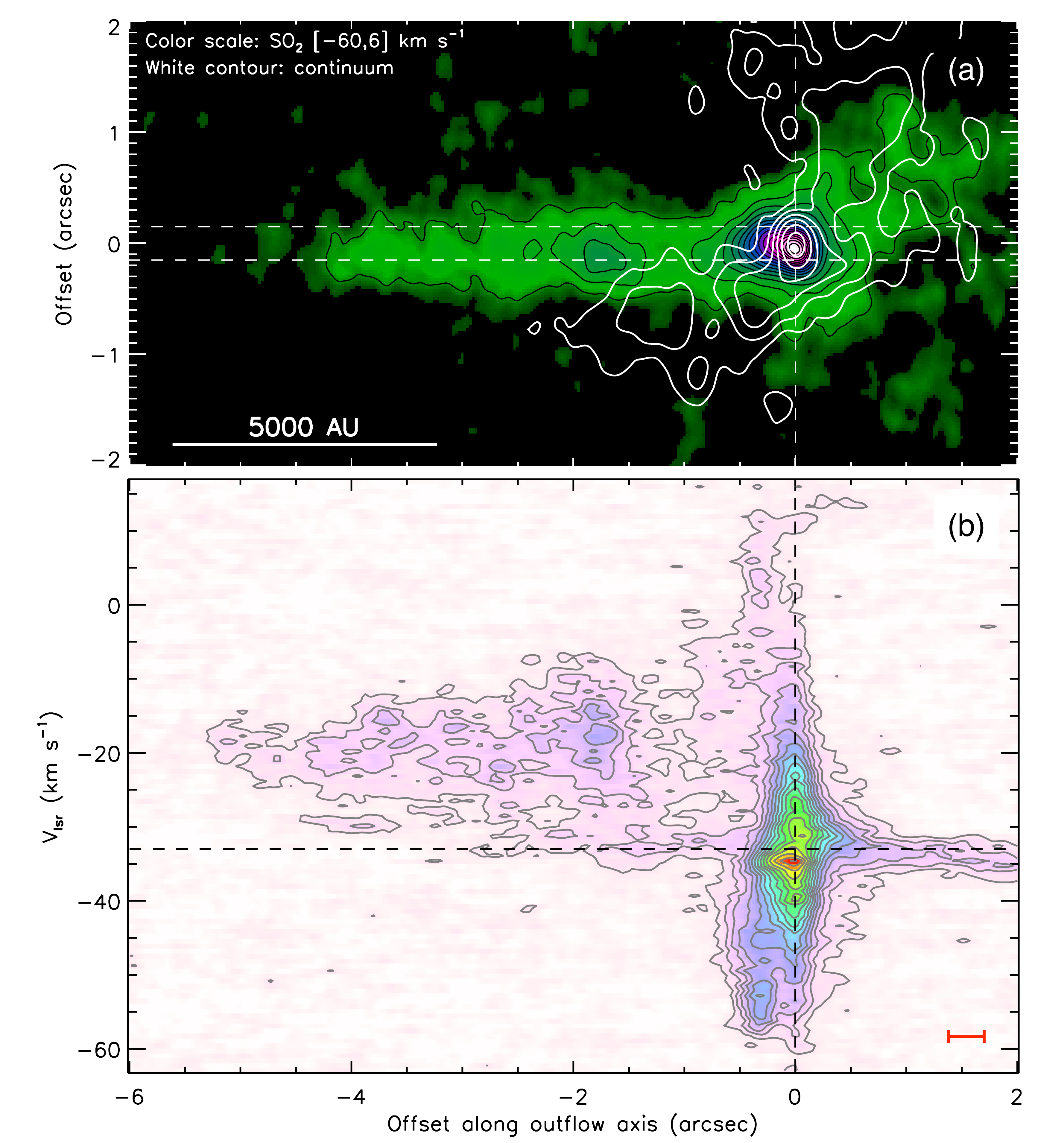}\\
\caption{
{\bf (a):} Integrated emission map of SiO($5-4$) emission (color
scale and black contours), overlaid with the continuum emission (white
contours). The images are rotated by $20^\circ$ clockwise so that the
SiO jet is along the $x-$axis. The SiO emission is integrated in the
range from $\vlsr=-60$ to $6~\kms$. The levels of the black and white
contours are same as those in Figure \ref{fig:intmap}(a). The
horizontal lines indicate the cut used to make the position-velocity
diagram in panel (b). The offsets are relative to the continuum peak
position.  {\bf (b):} The position-velocity diagrams of the SiO
emission along the jet axis.  The contours start from $5\sigma$ and
have intervals of $5\sigma$ ($1\sigma=2.3~\mJybeam$).  The red bar at
the lower-right corner indicates the resolution beam size.}
\label{fig:pvdiagram1}
\end{center}
\end{figure*}

\begin{figure}
\begin{center}
\includegraphics[width=\columnwidth]{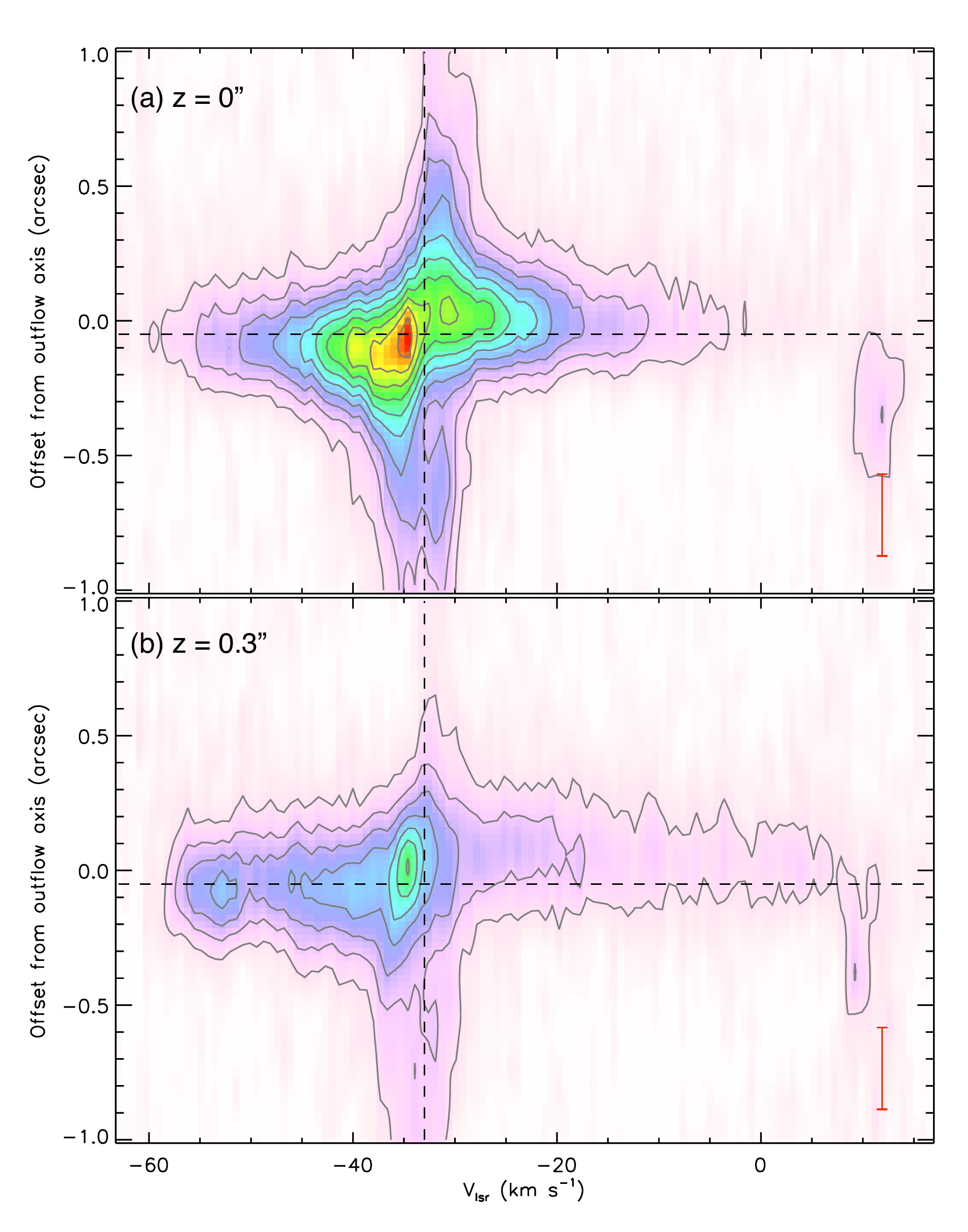}\\
\caption{Position-velocity diagrams of the SiO($5-4$) emission along two
cuts perpendicular to the SiO jet direction ($\pa=20^\circ$). The
distances of these two cuts to the central source are $0\arcsec$
(i.e., along the disk mid-plane; panel a) and $0.3\arcsec$ (630 au above
the disk mid-plane; panel b). The red bar at the lower-right corner of each
panel indicates the resolution beam size.}
\label{fig:pvdiagram2}
\end{center}
\end{figure}

Figure \ref{fig:pvdiagram1} shows a zoom-in view of the SiO jet and
the position-velocity diagram of the SiO emission along the jet
direction ($\pa=110^\circ$).  Only the red-shifted SiO jet is
detected. The average velocity of the SiO jet is about
$\vout=15\:\kms$.  Based on the spatial and velocity distribution of
the SiO emission, the SiO jet can be divided into two parts. The
first part is the extended structure starting from about $1\arcsec$
from the central source up to the end of the jet.  It is composed of
several knots, with the brightest one located at about $1.8\arcsec$
from the central source.  There are widespread velocities associated
with these knots, which is typical for the SiO emission from shocks
along the jet pathway. 

The second component is within about $1\arcsec$ from the central
source.  As discussed in \S\ref{sec:rotation}, the SiO emission at the
continuum peak position is dominated by the disk and inner
envelope, but the SiO emission peak is elongated up to $0.5\arcsec$
from the central source on the east side. As Figure
\ref{fig:pvdiagram1}(b) shows, this elongated structure has a velocity
range even wider than that at the central source position, from about
$\vlsr=-60$ to about $10~\kms$.  It has a strong blue-shifted emission
up to about $\vlsr=-60~\kms$, despite that the rest of the SiO jet is
red-shifted.  As discussed in \S\ref{sec:outflowtracer}, the outflow
is likely to be close to an edge-on view, i.e., lies in the plane of
the sky. In such a case, small changes in jet direction may cause a
change from red-shifted to blue-shifted, i.e., from being inclined
away from the observer to being inclined toward the observer. This
blue-shifted emission may represent the inclination of the most recent
jet activity. As discussed in \S\ref{sec:model}, this inclination is
consistent with how the envelope is inclined with respect to
the plane of the sky.

Another possibility is that these emissions at $0.3\arcsec$ from the
central source trace the material directly launched from the disk, and
the wide velocity range is due to the rotation of this disk wind.  The
SiO molecules in the jet structure can form via two different
mechanisms.  The first is that they are released to gas by dust
sputtering or grain-grain collisions in shocks during the interaction
between the jet and the surrounding material, or the interaction
between different components in the jet.  In such a case, the SiO
emission marks the location of strong shocks along the jet pathway.
The second mechanism is that the SiO molecules are released to the
gas phase in the disk by shocks or strong radiation, and then launched
to the outflow. In such a case, the SiO emission may trace the motion
of the gas that is directly launched from the disk, i.e., a disk wind.
The SiO emission tracing the directly launched material from the disk
is expected only in regions very close to the central source
(\citealt[]{Lee17,Hirota17}). In this source, since there is already
strong SiO emission associated with the disk, it is natural that some
SiO emission also traces the wind launched from the disk.

In order to better explore such a possibility, we compare the PV
diagrams of the SiO emission along two cuts perpendicular to the jet
direction in Figure \ref{fig:pvdiagram2}.  The distances of these
two cuts to the central source are $0\arcsec$ (along the disk
mid-plane, i.e., same as Figure \ref{fig:pvdiagram}a) and $0.3\arcsec$
(630~au above the disk mid-plane).  
Along the disk mid-plane, the emission velocity
ranges from $\vlsr=-60~\kms$ to $-3~\kms$, which is symmetric with
respect to the source velocity of $\vsys=-33~\kms$.  At $0.3\arcsec$
above the disk mid-plane, the SiO emission velocity ranges from
$\vlsr=-58~\kms$ to $7~\kms$, leading to a center velocity of
$\vlsr=-25~\kms$, i.e., $\vout=\vlsr-\vsys=12~\kms$, which is similar
to that of the other parts of the jet. This may suggest that at
0.3$\arcsec$ above the disk plane, the material has been accelerated
to outflowing velocities.  At this height, the PV diagram (Figure
\ref{fig:pvdiagram2}b) also shows a small level of velocity gradient
across the jet, with the most blue-shifted emission slightly to the
south and the most red-shifted emission slightly to the north, which
is consistent with the rotation direction in the disk (panel a).
Therefore, it is possible that the widespread velocities at
$0.3\arcsec$ above the disk mid-plane are due to the rotation in the
launched material.  The maximum rotation velocity detected at
$0.3\arcsec$ above the disk mid-plane is about $\vrot=33~\kms$, which
is higher than the maximum rotation velocity detected along the disk
mid-plane, which is about $\vrot=27~\kms$.  This is possible
considering that the specific angular momentum in the launched
material is a factor of several times that of material at the
launching point on the disk, according to magneto-centrifugal outflow
models (e.g., \citealt[]{Ferreira06}) and observations (e.g.,
\citealt[]{Zhang18}) in low-mass star formation.  Transition from disk
rotation to outflow rotation in massive protostellar sources has also
been inferred in Orion Source I (\citealt[]{Hirota17}).  Higher
resolution observations are needed to further confirm such a picture
in G339.

\subsection{Properties of the Large-scale CO Outflow}
\label{sec:largeoutflow}

\begin{table*} 
\scriptsize
\begin{center}
\caption{Derived properties of the CO outflow \label{tab:outflow}}
\begin{tabular}{lcccccc}
\hline
\hline
\multirow{2}{*}{Lobe$^\mathrm{a}$} & \multicolumn{3}{c}{Mass$^\mathrm{b,c}$ ($M_\odot$)} & \multicolumn{3}{c}{Momentum$^\mathrm{b}$ ($M_\odot~\kms$)}\\
\cline{2-7}
& Blue-shifted$^\mathrm{d}$ & Red-shifted$^\mathrm{d}$ & Total & Blue-shifted & Red-shifted & Total \\
\hline
Eastern & 0.07 & 0.13 & 0.20 & 0.72 & 1.3 & 2.0 \\
Western & 0.23 & 0.033 & 0.26 & 3.6 & 0.28 & 3.9 \\
Total & 0.30 & 0.16 & 0.46 & 4.3 & 1.6 & 5.9 \\
\hline
\end{tabular}
\end{center}
Notes:\\
$^\mathrm{a}$ The eastern and western lobes are divided by a line passing through the
central source with a position angle of $\pa=30^\circ$.\\ 
$^\mathrm{b}$ Assuming $\tex=17.5~\K$ (see text) and optically thin emission.\\
$^\mathrm{c}$ Not corrected for the inclination.\\
$^\mathrm{d}$ $\vlsr-\vsys<-5~\kms$ for blue-shifted and $\vlsr-\vsys>+5~\kms$ for red-shifted.\\
\end{table*}

\begin{figure}
\begin{center}
\includegraphics[width=\columnwidth]{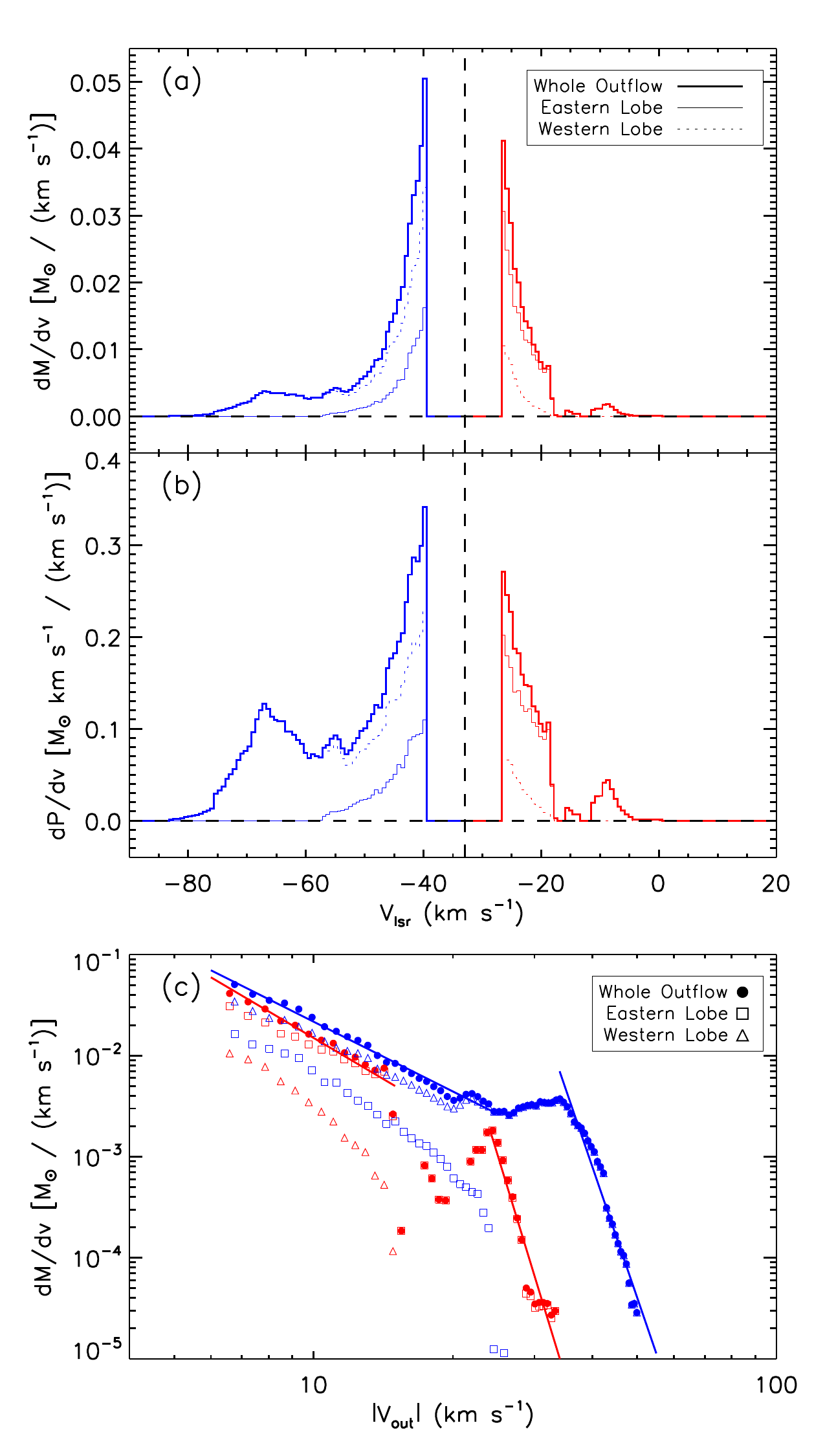}\\
\caption{
{\bf (a):} Distribution of outflow mass with channel velocity.  The
thick solid curves are for the whole outflow, the thin solid curves
are for the eastern outflow lobe, and the dotted curves are for the
western outflow lobe.  {\bf (b):} Distribution of outflow momentum
with channel velocity.  The momentum is not corrected for
inclination.  The line meanings are same as those in panel b.  {\bf
  (c):} The mass spectra of the blue-shifted (blue symbols) and
red-shifted (red symbols) outflows with outflow velocity
$\vout=|\vlsr-\vsys|$.  The circles are for the whole outflow, the
squares are for the eastern outflow lobe, and the triangles are for
the western outflow lobe.  The blue and red lines are the power-law
fits to the mass spectra of the total blue and red-shifted
outflows. The high-velocity parts of the mass spectra ($\vout>
34~\kms$ for the blue-shifted outflow, and $\vout>24~\kms$ for the
red-shifted outflow) and the low-velocity parts of the mass spectra
($\vout< 25~\kms$ for the blue-shifted outflow, and $\vout<15~\kms$
for the red-shifted outflow) are fitted separately.}
\label{fig:massspec}
\end{center}
\end{figure}

Figure \ref{fig:massspec} shows the mass and momentum distributions of
the outflow measured from the $^{12}$CO($2-1$) emission.  To obtain
the gas mass, we assume optically thin emission, and adopt an
abundance of $^{12}$CO of $10^{-4}$ relative to H$_2$ and a gas mass
of $2.34\times10^{-24}$ g per H$_2$ molecule.  
An excitation temperature of $\tex\approx 10 - 50~\K$ is typically used
for deriving the mass from low-$J$ CO transitions (\citealt[]{Dunham14}).
Within this range, an excitation temperature of $\tex=17.5~\K$ minimizes the mass
estimate from the CO($2-1$) line. 
An excitation temperature of 50~K would increase the mass
estimate (and therefore momentum estimate) by a factor of 1.5.
In each
velocity channel, we only include the primary beam corrected emissions
above $3\sigma$ within the region with the primary beam response
$>0.2$.  We exclude the emissions which are not related to the main
outflow on the east-west direction by applying masks that vary with
velocity channels.  We also exclude any emissions at velocities
$|\vlsr-\vsys|<5~\kms$ due to the confusion of the low-velocity
outflow emission with the core material emission.  Since the blue and
red-shifted emissions appear in both the eastern and western outflow
lobes, we also divide the two lobes by a line passing through the
central source with a position angle of $\pa=30^\circ$ (i.e.,
perpendicular to the direction of the large scale outflow; see
\S\ref{sec:outflowtracer}), and show the blue and red-shifted masses
of the individual lobes.

With $\tex=17.5~\K$, there are masses of 0.20 and 0.26~$M_\odot$ and
momenta of 2.0 and 3.9~$M_\odot~\kms$ in the eastern and western
outflow lobes, respectively.  In the eastern lobe, there are masses of
0.07 and 0.13~$M_\odot$ and momenta of 0.72 and 1.3~$M_\odot~\kms$ in
the blue and red-shifted materials, respectively.  In the western
lobe, there are masses of 0.23 and 0.033~$M_\odot$ and momenta of 3.6
and 0.28~$M_\odot~\kms$ in the blue and red-shifted materials,
respectively.  In total, there are masses of 0.30 and 0.16~$M_\odot$
and momenta of 4.3 and 1.6~$M_\odot~\kms$ in the blue and red-shifted
materials, respectively.
The whole outflow has a total mass of $m_w=0.46~M_\odot$ and a total
momentum of $p_w=5.9~M_\odot~\kms$.
The derived parameters are summarized in Table \ref{tab:outflow}.

The mass-weighted mean velocity of the outflow ($v_w=p_w/m_w$, where
$p_w=5.9~M_\odot~\kms$ and $m_w=0.46~M_\odot$
are the total outflow momentum and mass) is $13~\kms$,
which gives a dynamical timescale of $t_\mathrm{dyn}=1.2\times
10^4~\yr$, assuming the length of the outflow is $15\arcsec$.  
The mass outflow rate is then $\dot{m}_w=m_w/t_\mathrm{dyn}=3.9\times
10^{-5}~M_\odot~\yr^{-1}$.  
The momentum injection rate of the outflow is
$\dot{p}_w=p_w/t_\mathrm{dyn}=4.9\times
10^{-4}~M_\odot~\kms~\yr^{-1}$.  

There are several factors affecting these estimates. 
First, a different excitation temperature will increase these
estimates, e.g., $\tex=50~\K$ would increase the mass
estimate (and therefore the other estimates) by a factor of 1.5. 
Second, the correction factors for inclination on the
momentum, dynamical timescale, mass outflow rate, and
momentum injection rate are
$1/\sin i$, $\sin i/\cos i$, $\cos i/\sin i$, and $\cos i/\sin^2 i$,
respectively, where $i$ is the inclination angle between the outflow
axis and plane of sky. Assuming an
inclination of $i=20^\circ$, which is likely to be an upper limit (see
\S\ref{sec:outflowtracer}), the correction factors for inclination on the
momentum, dynamical timescale, mass outflow rate, and
momentum injection rate are 3.0, 0.36, 2.7 and 8.0, respectively.

Third, the optical depth effect and
the missing low-velocity outflow emissions will further increase the
mass and momentum estimates.  In one example of a low-mass
protostellar outflow, this factor is about 10 for mass estimation and
about 8 for momentum estimation (\citealt[]{Zhang16}).
Adopting these correction factors for the mass and momentum estimation,
and combining the correction factors for the inclination,
the total correction factors for the mass, momentum, dynamical timescale,
mass outflow rate, and momentum injection rate are
10, 24, 0.29, 34, and 82, respectively, which gives
$m_{w,\mathrm{cor}}=4.6~M_\odot$,
$p_{w,\mathrm{cor}}=1.4\times10^2~M_\odot~\kms$,
$t_\mathrm{dyn,cor}=3.5\times10^3~\yr$,
$\dot{m}_{w,\mathrm{cor}}=1.3\times10^{-3}~M_\odot~\yr^{-1}$,
$\dot{p}_{w,\mathrm{cor}}=4.0\times10^{-2}~M_\odot~\kms~\yr^{-1}$.

We note that the estimated outflow timescale is
likely to be at least an order of magnitude smaller than the formation
time of the protostar in Turbulent Core model of \citet[]{MT03},
so that only a part of the outflow history is being traced by these
observations. The derived mass and
momentum rates are consistent with the outflows of other protostellar
sources with total luminosities of several $\times 10^4~L_\odot$
(e.g., \citealt[]{Beuther02,Zhang05,Maud15,Yang18}).

Figure \ref{fig:massspec}(c) shows the mass spectra of the outflow.
The mass spectra can be characterized by two power laws, with sudden
changes of the spectrum slopes at $\vout=34~\kms$ in the blue-shifted
outflow and at $\vout=24~\kms$ in the red-shifted outflow. The
low-velocity mass spectra have power-law indices of $-2.3$ and $-2.7$
for the blue and red-shifted outflows, respectively, and the
high-velocity mass spectra have power-law indices of $-13.4$ and
$-15.0$ for the blue and red-shifted outflows, respectively.  The
break in the mass spectrum slope around $20-30~\kms$ may be related to
molecular dissociation caused by jet shocks (\citealt[]{Downes07}).
Such a cut-off at high velocities is also seen in some other massive
protostellar outflows, such as the G028.37+00.07 C1-Sa outflow
(\citealt[]{Tan16}), which also has the cut-off at
$\vout\approx20-30~\kms$.  The low-velocity mass spectrum slope index
around $-2.5$ is consistent with other massive outflows
(e.g., \citealt[]{Maud15}) and also low-mass outflows
(e.g., \citealt[]{Richer00}).  The low-velocity mass spectra may be
steeper if optical depth effects are considered.

\section{Discussion}
\label{sec:discussion}

\subsection{Origin of the Chemical Change across the Centrifugal Barrier}
\label{sec:chemistry}

The change of the type of molecular line emissions across the centrifugal barrier is
possible from a chemical point of view.  The CH$_3$OH and H$_2$CO
molecules are released from icy grain mantles to gas phase in the 
warm envelope, and strongly
enhanced around the centrifugal barrier (e.g. \citealt[]{Aota15}). 
The enhancement around the
centrifugal barrier may be due to the accretion shock, but more
importantly, may be due to the broadened inner edge of the
infalling-rotating envelope which can be directly irradiated by the
central source (e.g., \citealt[]{Sakai17}).  They may be destroyed
during the accretion shock at the same time, or reduced as they slowly
move inward in the disk.  There is also a lack of continuous supply of
CH$_3$OH and H$_2$CO molecules to the gas phase in the disk, since
they are mostly formed on the ice mantle of dust grains, 
and have already released into the gas phase around the centrifugal barrier.
Therefore the
CH$_3$OH and H$_2$CO emissions trace the envelope, have their emission
peaks around the centrifugal barrier, and do not show high rotation
velocities associated with the disk. 
In a similar manner, the SO$_2$
and H$_2$S molecules are also enhanced around the centrifugal barrier,
and gradually reduced as material moves toward the inner part of the
disk.  However, the observed transitions of these molecules have
higher upper energy levels than the CH$_3$OH and H$_2$CO transitions
(see Table \ref{tab:lines}),
so that they tend to trace the region further inward of the CH$_3$OH
and H$_2$CO emissions, i.e., the outer part of the disk inside of the
centrifugal barrier.  On the other hand, the accretion shock, internal
shocks, or a strong radiation field may have destroyed some fraction
of the dust grains to liberate SiO, which can remain in the gas phase
in the disk (or disk surface). Meanwhile, destruction of dust grains
in the disk provides a continuous supply of SiO molecules.  Therefore
the SiO emission can reach higher rotational velocities in the inner
disk.

As discussed in \S\ref{sec:envelopetracer}, the CH$_3$OH and H$_2$CO
emission is also enhanced along the outflow cavity.  In fact, Figure \ref{fig:MIR}(a) 
shows that the locations of the methanol masers in
this source coincide very well with the positions of strong thermal methanol
emissions, which is to the south of the central source and elongated
along the outflow direction.
The methanol masers have velocities around $\vlsr=-39$ to $-34~\kms$,
which are also consistent with the elongated structure in the outflow direction shown in the
thermal methanol emissions (see Figure \ref{fig:momentmap}c and 
Appendix Figure \ref{fig:chanmap_CH3OH}).
It is possible that the shocks due to
the outflow have enhanced the thermal emission and also excite the
maser emission of methanol along the outflow cavity wall.
Note that the transition region between the envelope and disk
is also expected to be the base of outflow cavity. 
Therefore, in this source, the thermal methanol emissions appear to trace both
the envelope (along the mid-plane) and the outflow cavity walls.
Higher angular resolution observations are needed to clearly separate
these components.

There is also an asymmetry in the distributions of the CH$_3$OH and H$_2$CO 
emissions, with the southern blue-shifted emissions much brighter than the northern
red-shifted emissions. 
One reason is that the northern red-shifted emission is affected 
by the self-absorption of the foreground infalling material in the envelope,
which has a slightly red-shifted velocity.
This effect is especially strong in the CH$_3$OH and H$_2$CO emissions as
they only trace the low-velocity components.
Similar behaviors have also been seen in low-mass sources for molecular lines tracing
the infalling-rotating envelope (e.g., \citealt[]{Sakai14b,Oya16}).
Another reason is that the shocks may be stronger on the southern side,
which enhance the CH$_3$OH and H$_2$CO emissions.
The CH$_3$OH maser emissions are also seen on the southern side (see Figure \ref{fig:MIR}a),
supporting this scenario. The asymmetric shock conditions may be a result of
a clumpy distribution of material in the infalling envelope.

\subsection{Connecting Multi-wavelength Observations from Mid-infrared to Radio}
\label{sec:multiwave}

\begin{figure*}
\begin{center}
\includegraphics[width=0.9\textwidth]{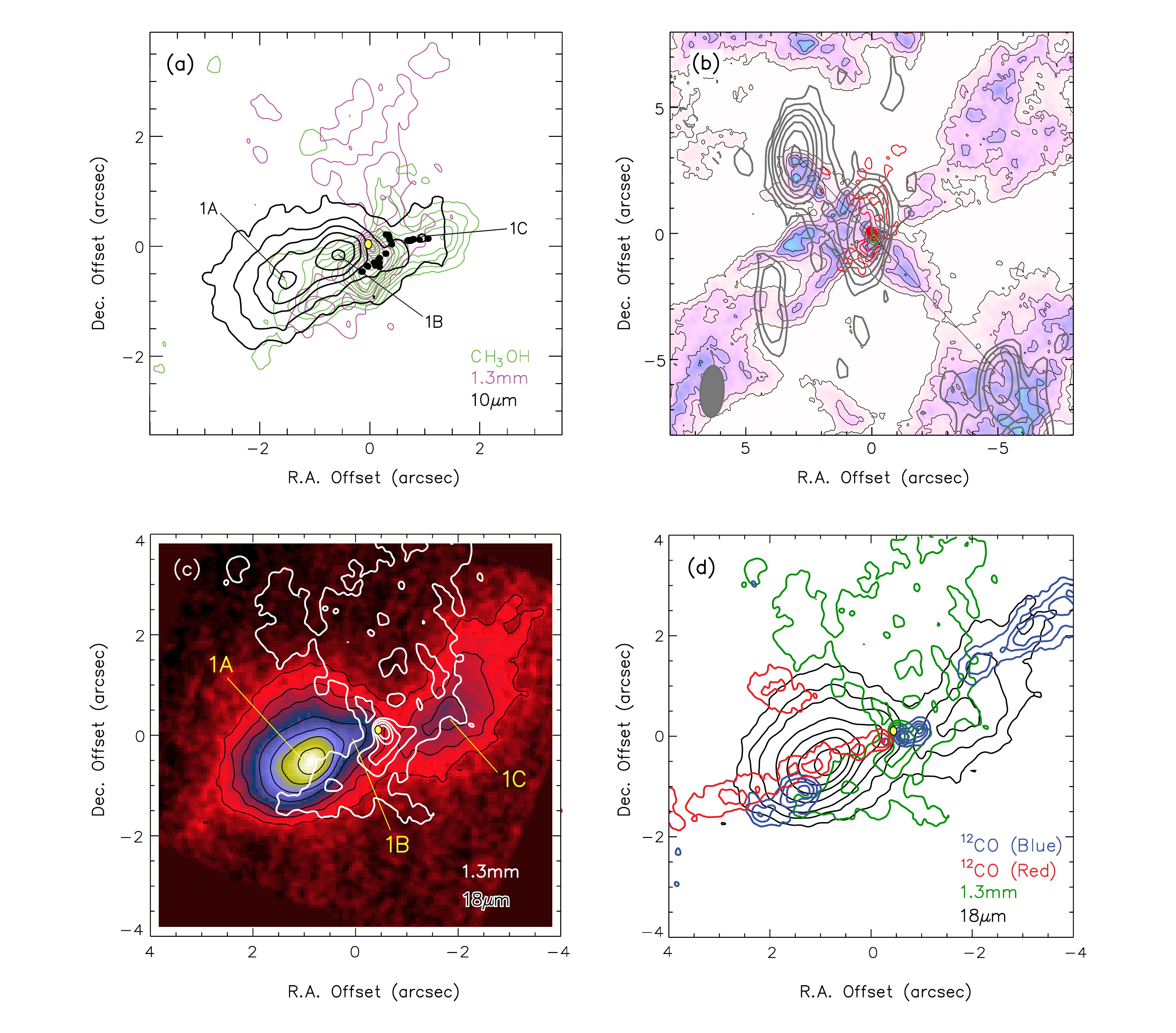}\\
\caption{{\bf (a):} The 1.3 mm continuum emission (purple contours)
and CH$_3$OH ($4_{2,2}-3_{1,2}$; E) integrated emission (green contours),
overlaid with the 10 $\mu$m continuum emission (black contours). 
The black dots are the locations of CH$_3$OH masers and
the yellow circle marks of peak position of the radio continuum emission.
{\bf (b):} The $^{12}$CO ($2-1$) emission at $\vlsr=-39~\kms$ (color-scale and contours)
overlaid with the 9 GHz radio continuum emission (black contours)
and 1.3 mm continuum emission (red contours).
{\bf (c):} The 18 $\mu$m  continuum emission map (color scale and contours) overlaid with
the 1.3 mm continuum emission (white contours).
{\bf (d):} The integrated $^{12}$CO ($2-1$) outflow emissions (blue and red contours)
overlaid with the 18 $\mu$m (black contours) and 1.3 mm continuum (green contours) emissions.
The MIR and maser maps are taken from \citet[]{Debuizer02}.
The 9 GHz radio continuum emission is taken from \citet[]{Purser16}.}
\label{fig:MIR}
\end{center}
\end{figure*}

Figure \ref{fig:MIR} compares the previously observed 10 $\mu$m, 18 $\mu$m, and
9 GHz continuum emissions (\citealt[]{Debuizer02,Purser16}),
with the ALMA 1.3 mm continuum and molecular line emissions.
The 1.3 mm continuum peak does not coincide with any of the three
peaks identified in the 10 $\mu$m emission (panel a), which
challenges the claim by \citet[]{Debuizer02} that the MIR peak 1B is
an embedded protostar driving the radio jet. 
The 9 GHz radio continuum emissions
reveal three knots, with the central knot lying very close to the
1.3 mm continuum source (panel b).
The ALMA observation shows $^{12}$CO emissions
at the positions of the northern and southern radio continuum lobes at velocities
close to the systemic velocity (panel b; see also Figure \ref{fig:chanmap_12CO}).
Other molecular species observed by ALMA, which trace higher densities
than $^{12}$CO, do not show emissions concentrated toward these northern
and southern lobes, unlike protostellar HC/UC HII regions
that are associated with dense molecular structures.
These features support a scenario in which the northern and southern radio continuum lobes 
are actually tracing an outflow from the main source, rather than being from separate massive protostars.

The extension of the MIR emission is in the same direction as the CO outflow (panel d),
suggesting that the MIR emissions are coming from the outflow and/or outflow cavity.
This is consistent with previous observations of other massive protostellar sources and
theoretical models, which have shown that the $10-20~\mu$m continuum
emissions are strongly affected by the outflow cavity structures in
massive young stellar objects (MYSOs) (e.g.,
\citealt[]{Debuizer06,Zhang13,Debuizer17,Zhang14}). 
The driving source of this outflow, however, is not located at
the MIR peak 1B as speculated previously. 
In fact, the 1.3 mm continuum source is located close to the gap seen in 
the 18 $\mu$m emission (panel c), which can be naturally explained by
the high extinction of the dust concentrated around the central source.
The 1.3 mm continuum peak also coincides with the peak of the dust color temperature distribution
derived from the 10 and 18 $\mu$m emissions (\citealt[]{Debuizer02}),
which was believed to indicate a massive star slightly in the foreground and less obscured
(see \S\ref{sec:target}).
The ALMA observation suggests that it is the heating from the embedded driving source that is responsible
for this temperature peak.

The prediction of MIR emission from outflow cavities is that the 
blue-shifted (near-facing) side should be brighter in the MIR as the red-shifted (far-facing) 
side is more obscured by the envelope (\citealt[]{Debuizer06,Zhang13,Zhang14}). 
However, we see the opposite things in the CO and MIR emissions for G339.
The eastern outflow, which is mostly red-shifted, has brighter MIR emission than 
the western outflow, which is mostly blue-shifted.
One possible reason for this is the almost edge-on view of this outflow. 
In such a case, the brightness distribution of the MIR emission is not dominated by
the overall inclination of the outflow cavity, but strongly affected by 
the detailed distribution of the extended cold dust in the region.
Figure \ref{fig:MIR}(c) shows that cold dust traced by mm continuum emission
is distributed north, south and west of 
the central source. If this dust is in the foreground of the MIR emission, extinction would explain why the 
MIR peak 1A (where there is no 1.3 mm emission) is so much brighter than source 1C 
(which is behind the extended mm emission). 
In fact, the MIR emission from 1A appears to be constricted 
by the extended mm emission from the south and north. 

To summarize, the ALMA observations of 1.3 mm continuum and molecular line emissions,
combined with the MIR and radio continuum observations, support the following scenario for G339.
The central source is likely to contain an unresolved massive protostellar system, 
which may be a binary, but is being fed in a relatively ordered way from the infall envelope. 
The main source drives an outflow in the east-west direction, seen in $^{12}$CO and MIR continuum.
A second outflow may be present in the northeast-southwest direction, seen in radio continuum and 
also $^{12}$CO emission. The MIR emission is dominated by the main outflow cavities, 
with its brightness distribution affected by the distribution of the extended cold dust.

\subsection{Implications for Massive Star Formation}
\label{sec:massivestarformation}

The estimated radius of the centrifugal barrier in G339
($\rcb=530\pm100~\au$) is similar to the radius of the centrifugal
barrier recently identified in the massive protostellar source
G328.2551-0.5321, which is $300-800~\au$ (\citealt[]{Csengeri18}).  In
massive source G17.64+0.16, indication of a change of kinematics from
rotation with radial motion to pure rotation is found at a radius of
$\sim 200~\au$, which may also suggest a centrifugal barrier at that
radius (\citealt[]{Maud18}).  If the centrifugal barrier is the outer
boundary of the disk, the measured radius of the centrifugal barrier in
G339 also indicates a smaller disk than most disks or circumstellar
structures identified around massive protostars, which typically have
sizes of several $\times 10^3~\au$ (\citealt[]{Beltran16}), although
most of these structures are likely to be pseudo disks, which are
expected to extend further out of the centrifugal barrier, rather than
Keplerian disks.  Still, the G339 disk is smaller than some recently
identified Keplerian disks around massive protostars, e.g.,
G35.2-0.74N ($R=2600~\au$; \citealt[]{Sanchez13}), AFGL 4176
($R=2000~\au$; \citealt[]{Johnston15}), G11.92-0.61 MM1 ($R=1200~\au$;
\citealt[]{Ilee16}).

On the other hand, the estimated radius of the centrifugal barrier in
G339 is larger than those found around low-mass protostars, which are
typically smaller than 200 au (e.g.,
\citealt[]{Sakai14,Oya16,Oya17,Alves17}).  The specific angular
momentum measured at the centrifugal barrier in G339 is
$\rcb\vcb\approx3000~\au~\kms$, which is about an order of magnitude
higher than those found in low-mass systems (e.g.,
\citealt[]{Sakai14,Oya16}). The total mass, momentum, and momentum
injection rate in the G339 outflow are also significantly higher than
those of typical low-mass protostellar outflows (e.g.,
\citealt[]{Dunham14,Zhang16}).  However, in spite of these
quantitative differences, G339 is found to be qualitatively very
similar to lower-mass protostars, including possessing a highly
collimated outflow, and ordered transition from an infalling-rotating
envelope to a Keplerian disk, which is also accompanied by 
change of types of molecular line emissions.

As mentioned above, our analysis is based on a simplified model in
which the effects of the magnetic field are not considered.  It is
likely that the magnetic field is playing at least two essential roles
in this source.  The first is magnetic braking, not only affecting
disk formation, but also suppressing the fragmentation of the core.
Despite the massive envelope containing hundreds of thermal Jeans
masses, the lack of fragmentation into small compact sources implies
efficient support against collapse and/or transportation of angular
momentum by magnetic braking (e.g.,
\citealt[]{Seifried11,Commercon11}).
The second role of the magnetic field is driving the
outflow. Recently, both observations (e.g., \citealt[]{Hirota17}) and
magneto-hydrodynamical (MHD) simulations (e.g.,
\citealt[]{Matsushita17,Staff18}) showed that magneto-centrifugal disk
winds can arise in massive star formation in a similar way as in
low-mass star formation.  The observed outflow motion in G339 is also
consistent with the magneto-centrifugal outflow (see
\S\ref{sec:launch}).  Higher resolution observations will provide more
detailed information about the processes of magnetic braking and
magneto-centrifugal wind launching as both are most efficient inside
the centrifugal barrier where the magnetic field becomes twisted and
tightly wrapped.

Recently, \citet[]{Liu19} compared MIR observations of this
source with the SED model grid of massive star formation
(\citealt[]{ZT18}).  The model grid is based on the Turbulent Core
model of massive star formation (\citealt[]{MT03}), and includes
evolutions under various initial and environmental conditions.  The
best-fit models also show narrow outflow cavities with half opening
angles of $10^\circ-20^\circ$, which is consistent with our
observations. Among the five best models, there is one model that has
a close edge-on view with inclination of $i\approx 20^\circ$ between
the outflow axis and the plane of sky. This suggests that the ALMA
observations of the outflow can help to break the degeneracies in the
SED fitting of infrared fluxes.  That model has a protostellar mass of
$12~M_\odot$, consistent with the dynamical mass estimated from the
gas kinematics. This particular model gives 
a total envelope mass of $\sim300~M_\odot$ within a radius of
$\sim 30\arcsec$, and an envelope mass of
about $17~M_\odot$ within 10,000 au. 
The latter is close to the total mass
measured from the continuum emission associated with the central
source, which is $23\:M_\odot$ assuming a dust temperature of 70~K (see
\S\ref{sec:continuum}).
The outflow mass estimated from
$^{12}$CO emissions is about 5~$M_\odot$ 
(within $\sim 15\arcsec$), including a factor of
$\sim10$ for corrections of optical depth and missing low-velocity
emissions (see \S\ref{sec:largeoutflow}).  If the majority of the
$^{12}$CO outflow is entrained material, and if the outflow
entrainment is the reason for outflow cavity widening, the mass in the
envelope should be about $\cos\theta_w/(1-\cos\theta_w)\approx 15$
times the outflow mass, i.e., $\sim 75~M_\odot$, where we assume an
half opening angle of the outflow of $\theta_w=20^\circ$, and
isotropic distribution of material in the core. 
This mass is somewhat larger, but still on similar levels as, the measured
envelope masses from the 1.3 mm continuum ($23~M_\odot$) and SED modeling ($17~M_\odot$),
considering that the outflow mass is measured on a larger scale ($\lesssim 15\arcsec$) 
than the 1.3 mm continuum ($\lesssim 10,000~\au$).
This result supports a scenario in which the outflow cavity opens up as material is
gradually entrained into the outflow (i.e., \citealt[]{Arce06}), and is also
expected in the outflow feedback model for massive star formation
based on the Turbulent Core Accretion model
(\citealt[]{Zhang14,Tanaka17,ZT18,Staff18}).

Overall, our results strongly indicate that, at least in this
particular example, massive star formation can be considered to be a
scaled-up version of low-mass star formation, i.e., forming via
relatively ordered core accretion. Furthermore, there is quantitative
consistency in many parameters of the system as derived by comparing to
semi-analytic models based on the Turbulent Core Accretion theory of
\citet{MT03}.  The larger centrifugal barrier radius and specific
angular momentum compared to lower-mass sources can be explained by
collapse of a rotating massive core on larger scales.  
On the other
hand, in Competitive Accretion, the
fragmentation of the gas clump into many small low-mass interacting
cores and protostars, including outflows, is likely to prevent the formation of an ordered
disk and rotating infall envelope on these $\sim500~\au$
scales. The disk sizes in the Competitive Accretion model are expected to be
much more compact than in Core Accretion from a massive core (e.g., \citealt[]{Kratter06}).
The main outflow seen in the $^{12}$CO emission is highly collimated
and extends to $>15\arcsec$ ($0.15~\pc$).
This suggests that the outflow
has not experienced significant disturbance from interactions with
other protostars in the region during the last $\sim10^4\:$yr, which is not expected in
Competitive Accretion models.
The lack of multiple compact continuum sources in the region,
i.e., an apparent relatively low multiplicity, also supports the Core
Accretion scenario rather than Competitive Accretion scenario.

\section{Summary}
\label{sec:summary}

We have presented ALMA observations of the envelope, disk, and outflow
system in the massive protostellar source G339.88-1.26 (G339),
including $^{12}$CO, SiO, C$^{18}$O, CH$_3$OH, H$_2$CO, SO$_2$, and
H$_2$S emissions. Our main conclusions are as follows.

1) The $^{12}$CO($2-1$) emission reveals a collimated bipolar outflow
driven by the G339 protostar.  The SiO($5-4$) emission also reveals
the red-shifted jet at the base of the large scale $^{12}$CO outflow.

2) The envelope/disk system is traced by SiO($5-4$),
SO$_2$($22_{2,20}-22_{1,21}$), H$_2$S($2_{2,0} - 2_{1,1}$),
CH$_3$OH($4_{2,2}-3_{1,2}$; E), and H$_2$CO($3_{2,1} - 2_{2,0}$)
emissions.  Based on their spatial distributions and kinematics, we
found that these molecular lines trace different parts of the
envelope-disk system.  The SiO emission traces the disk and inner
envelope. The CH$_3$OH and H$_2$CO emissions trace the
infalling-rotating envelope outside of the disk. The SO$_2$ and H$_2$S
emissions appear to be enhanced around the transition region between
the envelope and disk, i.e., the centrifugal barrier, and trace the
outer part of the disk. Therefore, the transition from an envelope to
a disk is not only seen in the change of kinematics, but also in the
change of types of molecular line emissions.

3) The kinematics of the envelope can be well fit by a model of
infalling-rotating motion.  In such a model, the envelope collapses
with the angular momentum conserved, and the kinetic energy is
completely converted to rotation at its inner boundary, i.e., the
centrifugal barrier.  Based on our model fitting, we estimate the
radius of the centrifugal barrier to be about $530\pm100~\au$ and the
rotation velocity at the centrifugal barrier to be about $6\pm1~\kms$,
leading to a central mass of about $11^{+6}_{-5}~M_\odot$. Inside of
the centrifugal barrier, the rotation appears to be consistent with
Keplerian rotation, but higher resolution observations are needed to
confirm it.

4) We found that the SiO emission slightly above the disk plane
($\sim0.3\arcsec$) may trace the outflowing material launched from the
disk. This emission shows signature of rotation, which smoothly
connects to emission in the disk mid-plane, indicating angular
momentum transfer from the disk to the outflow.

5) We estimate a total mass of 0.46 $M_\odot$ and a total momentum of
$5.9~M_\odot~\kms$ in the outflow based on the $^{12}$CO emission,
without any corrections for inclination or optical depth.  
After correcting for these effects, we estimate, very roughly, that the total mass, momentum, mass outflow rate
and momentum injection rate are
$4.6~M_\odot$,
$1.4\times10^2~M_\odot~\kms$,
$1.3\times10^{-3}~M_\odot~\yr^{-1}$, and
$4.0\times10^{-2}~M_\odot~\kms~\yr^{-1}$, respectively.

6) Combined with previous MIR and radio continuum observations, 
the ALMA observations suggest that 
the central source drives an outflow in the east-west direction, 
seen in $^{12}$CO and MIR continuum.
A second outflow may be present in the northeast-southwest direction, seen in radio continuum and 
also $^{12}$CO emission. The MIR emission is dominated by the main outflow cavities, 
with its brightness distribution affected by the distribution of the extended cold dust.

7) The envelope-disk-outflow system detected in the massive
protostellar source G339 appears to be highly ordered and
qualitatively very similar to those observed around low-mass
protostars, even though the disk size, angular momentum and outflow
strength are much larger than in the lower-mass cases. Our results
imply that at least some massive stars form in a way that is similar
to those of low-mass stars, i.e., via Core Accretion.

\acknowledgements
This paper makes use of the following ALMA data:
ADS/JAO.ALMA\#2015.1.01454.S.  ALMA is a partnership of ESO
(representing its member states), NSF (USA) and NINS (Japan), together
with NRC (Canada), MOST and ASIAA (Taiwan), and KASI (Republic of
Korea), in cooperation with the Republic of Chile.  The Joint ALMA
Observatory is operated by ESO, AUI/NRAO and NAOJ.  Y.Z. acknowledges
support from RIKEN Special Postdoctoral Researcher
Program. J.C.T. acknowledges support from NSF grant AST1411527 and ERC
project 788829 - MSTAR. 
N. S. acknowledges support from JSPS KAKENHI grant 18H05222.
K.E.I.T. acknowledges support from NAOJ ALMA
Scientific Research Grant Number 2017-05A.
D.M. and G.G. acknowledge support from CONICYT project AFB-170002

\software{CASA (http://casa.nrao.edu; \citealt[]{McMullin07}), 
The IDL Astronomy User's Library \\(https://idlastro.gsfc.nasa.gov)}

\appendix
\section{Channel Maps of the Molecular Lines}
\label{sec:appA}

\begin{figure}
\begin{center}
\includegraphics[width=\textwidth]{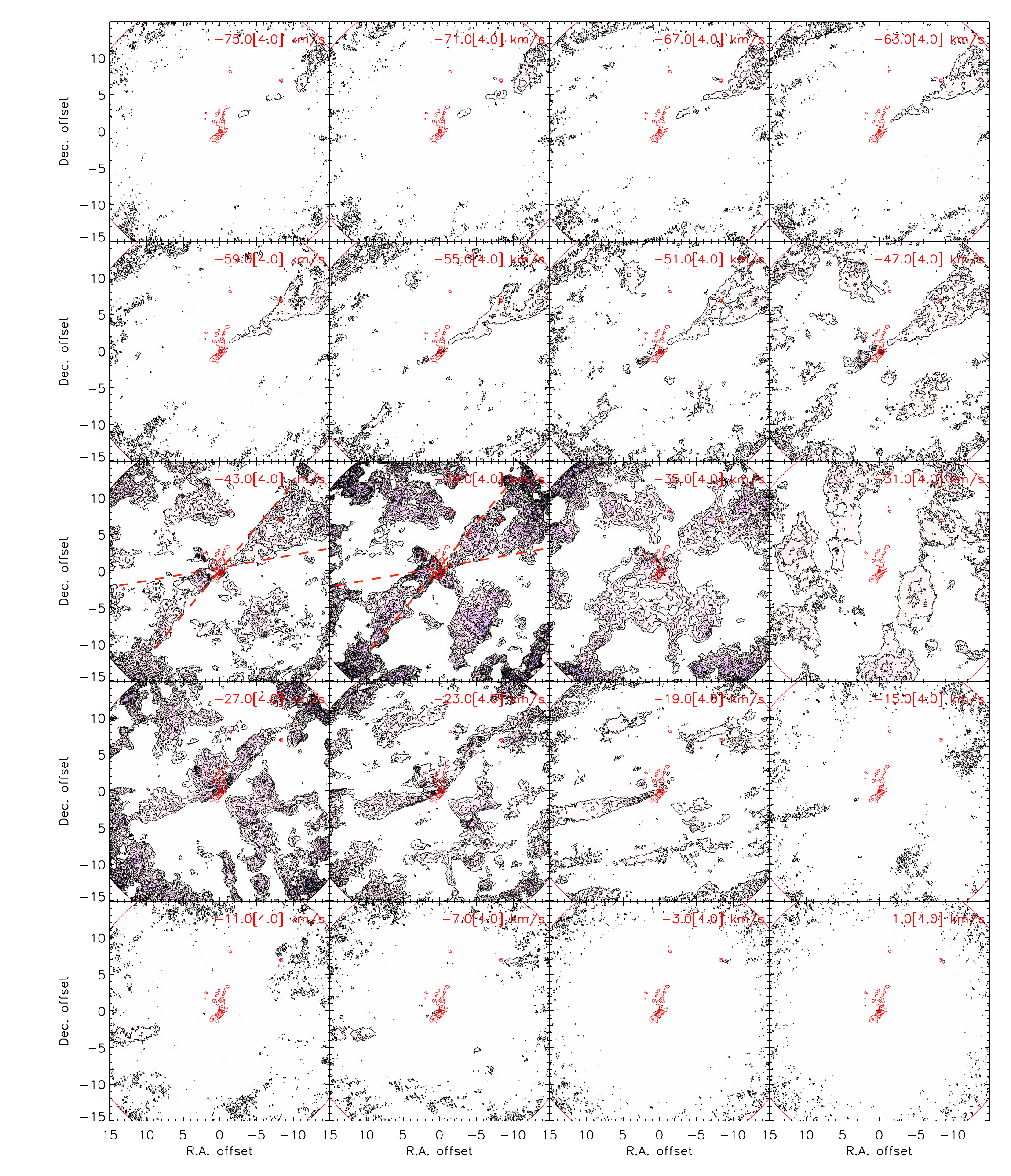}\\
\caption{
Channel maps of the $^{12}$CO($2-1$) emission (color scale and
black contours).  The central velocity and the width of each channel
are labeled on each panel. The black contours start at $5\sigma$
and have intervals of $10\sigma$ ($1\sigma=2.5~\mJybeam$).
The red contours show the continuum emission.
The red dashed lines show the opening angle of the main outflow.}
\label{fig:chanmap_12CO}
\end{center}
\end{figure}

\begin{figure}
\begin{center}
\includegraphics[width=\textwidth]{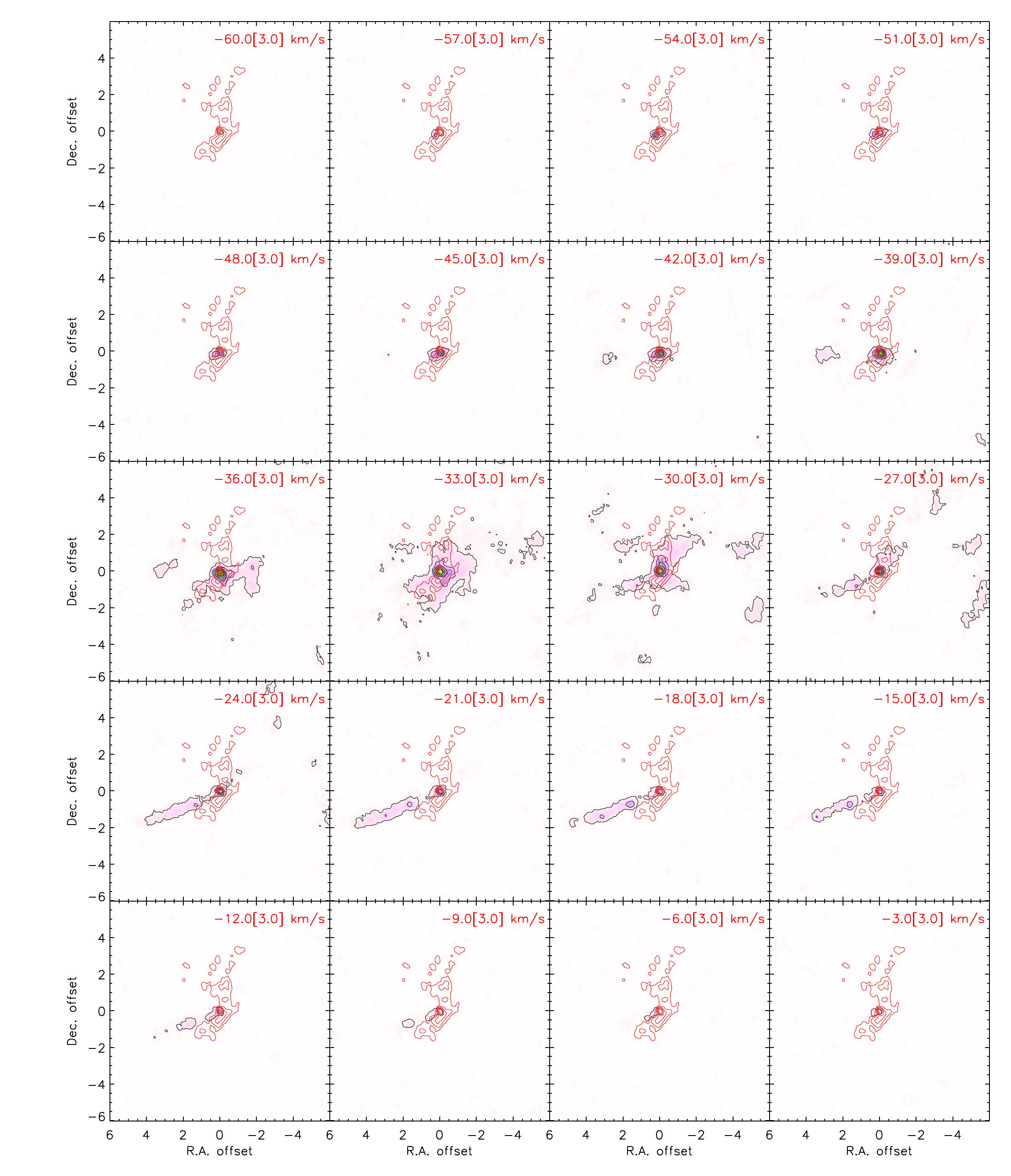}\\
\caption{
Same as Figure \ref{fig:chanmap_12CO}, but for the SiO$(5-4)$
emission. The black contours start at $5\sigma$
and have intervals of $10\sigma$ ($1\sigma=1.9~\mJybeam$).}
\label{fig:chanmap_SiO}
\end{center}
\end{figure}

\begin{figure}
\begin{center}
\includegraphics[width=\textwidth]{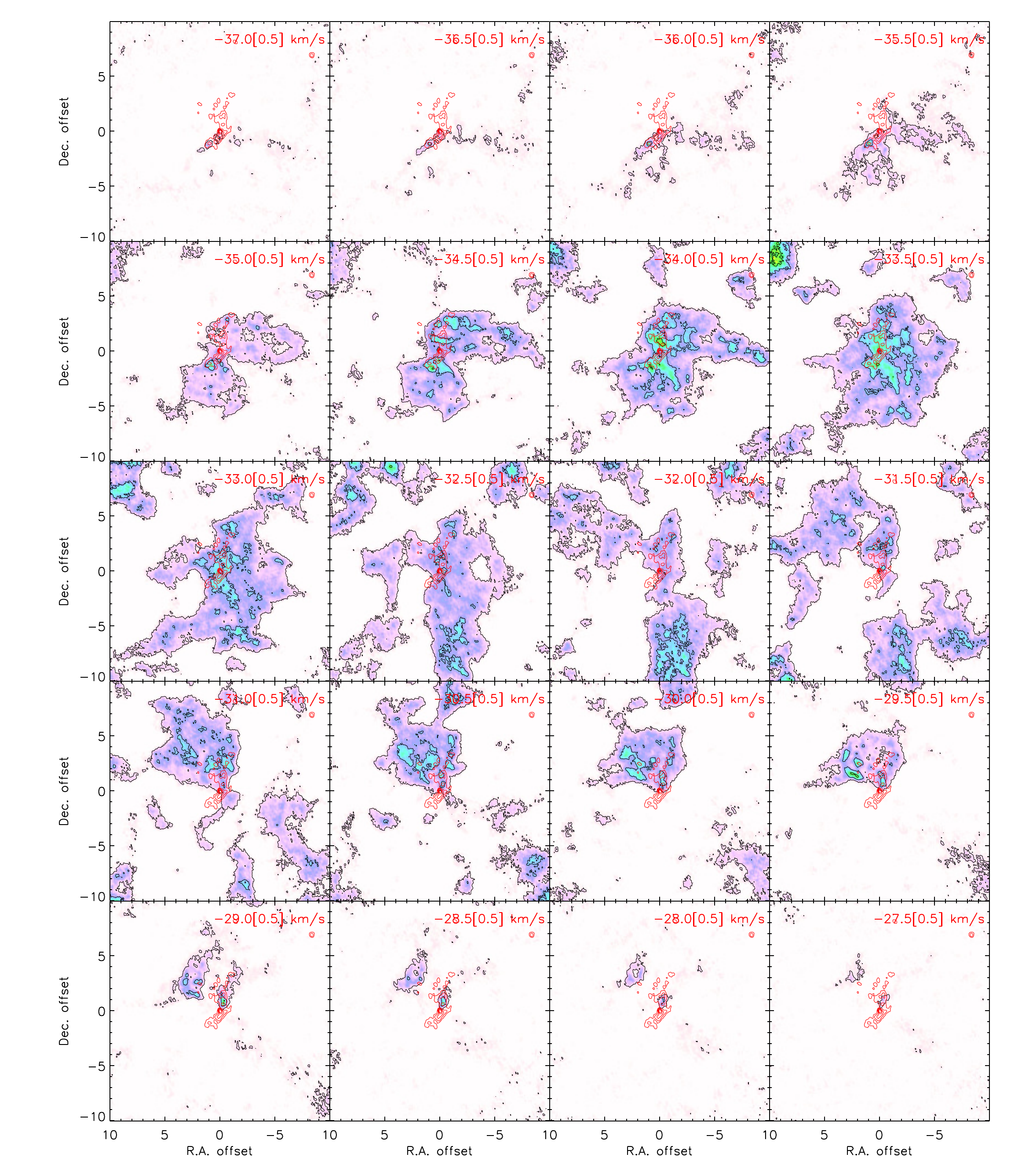}\\
\caption{Same as Figure \ref{fig:chanmap_12CO}, but for
the C$^{18}$O$(2-1)$ emission. The black contours start at $5\sigma$
and have intervals of $10\sigma$ ($1\sigma=4.4~\mJybeam$).}
\label{fig:chanmap_C18O}
\end{center}
\end{figure}

\begin{figure}
\begin{center}
\includegraphics[width=\textwidth]{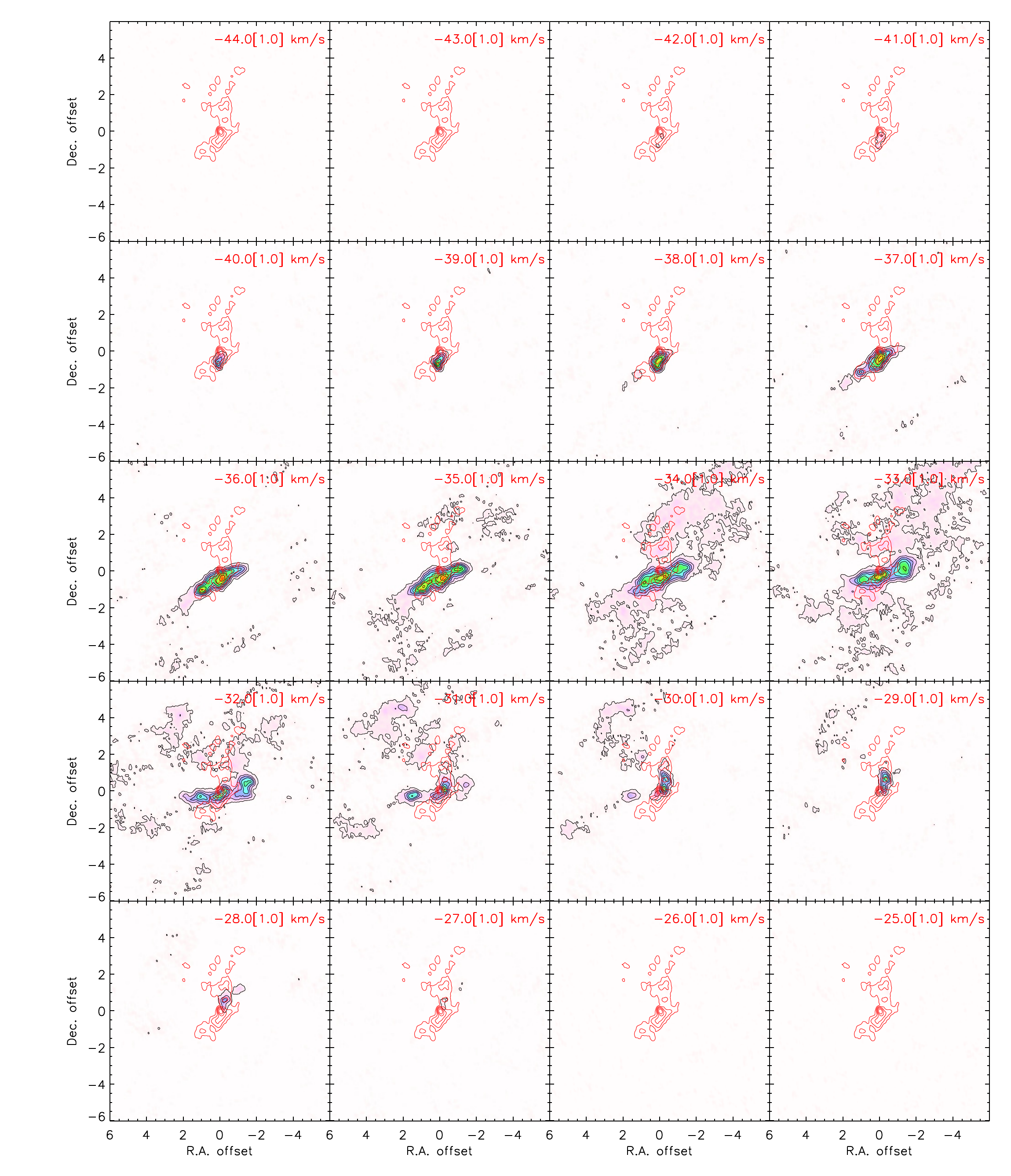}\\
\caption{Same as Figure \ref{fig:chanmap_12CO}, but for
the CH$_{3}$OH($4_{2,2}-3_{1,2}$; E) emission. The black contours start at $5\sigma$
and have intervals of $10\sigma$ ($1\sigma=3.5~\mJybeam$).}
\label{fig:chanmap_CH3OH}
\end{center}
\end{figure}

\begin{figure}
\begin{center}
\includegraphics[width=\textwidth]{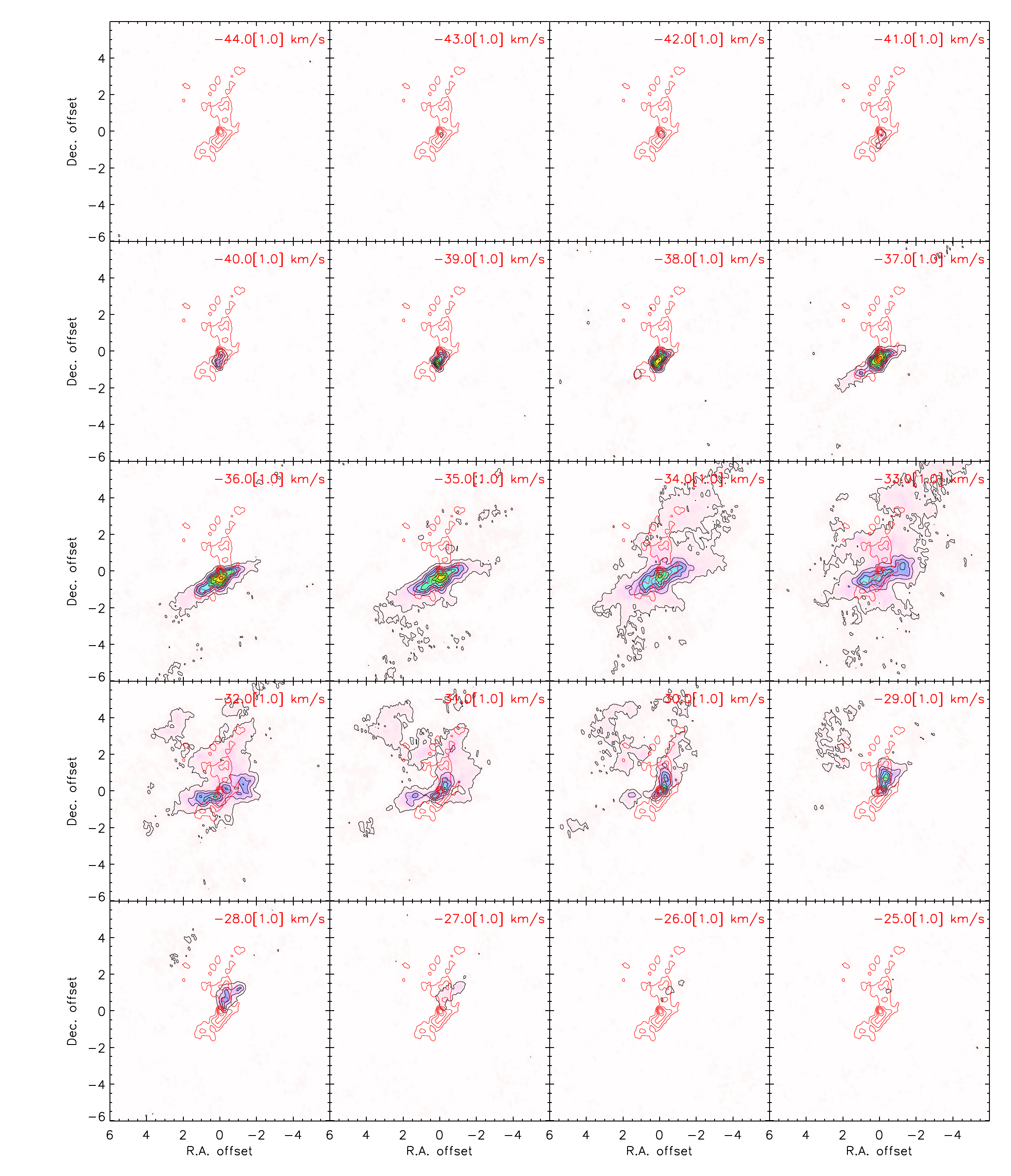}\\
\caption{Same as Figure \ref{fig:chanmap_12CO}, but for
the H$_2$CO($3_{2,1} - 2_{2,0}$) emission. The black contours start at $5\sigma$
and have intervals of $10\sigma$ ($1\sigma=3.0~\mJybeam$).}
\label{fig:chanmap_H2CO}
\end{center}
\end{figure}

\begin{figure}
\begin{center}
\includegraphics[width=\textwidth]{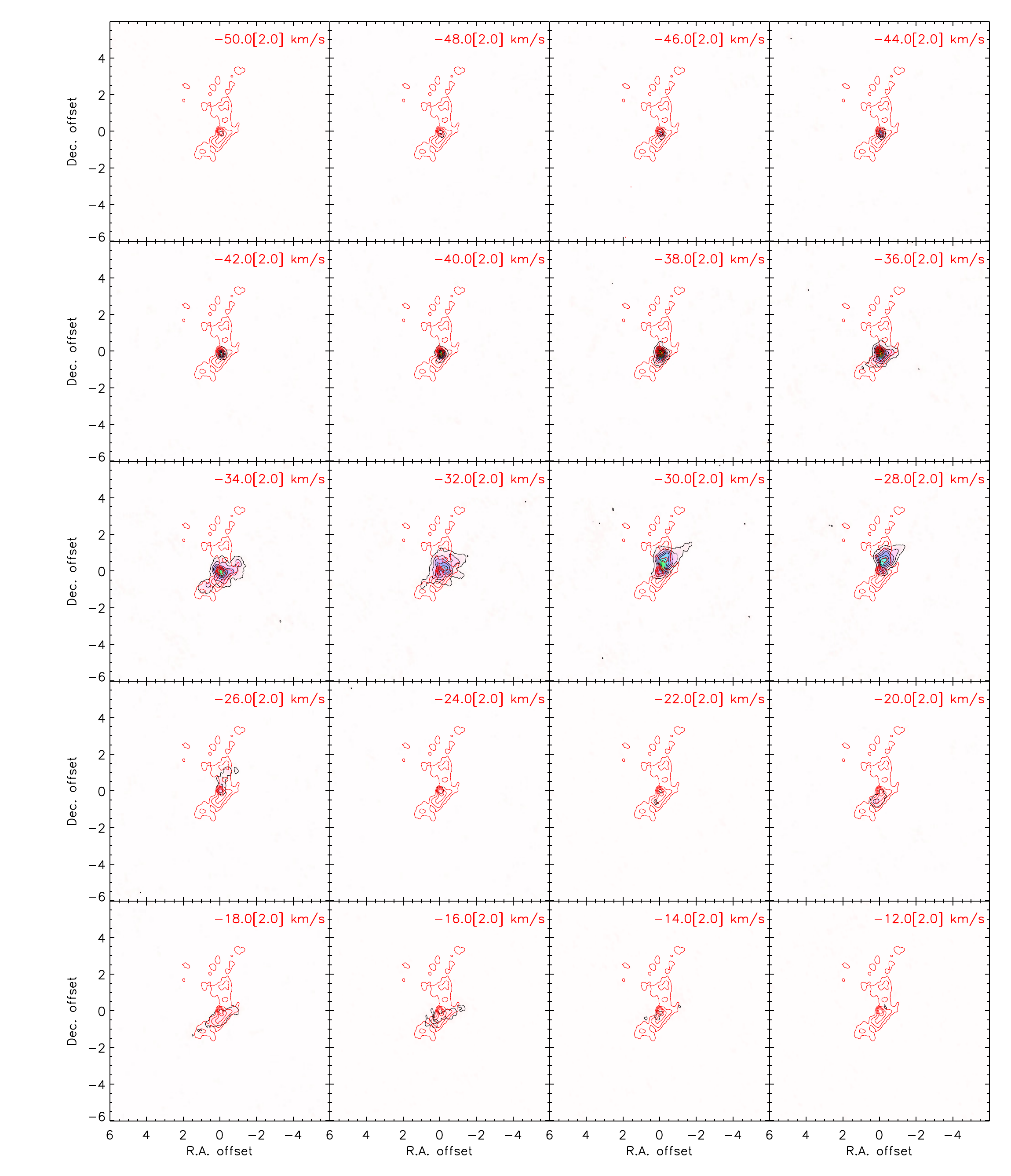}\\
\caption{
Same as Figure \ref{fig:chanmap_12CO}, but for the
SO$_2$($22_{2,20}-22_{1,21}$) emission. The black contours start at $5\sigma$
and have intervals of $10\sigma$ ($1\sigma=2.2~\mJybeam$).
The emissions at velocities of $\vlsr=-20$ to $-14~\kms$ are from 
the CH$_3$CHO ($11_{1,10}-10_{1,9}$; E) transition.}
\label{fig:chanmap_SO2}
\end{center}
\end{figure}

\begin{figure}
\begin{center}
\includegraphics[width=\textwidth]{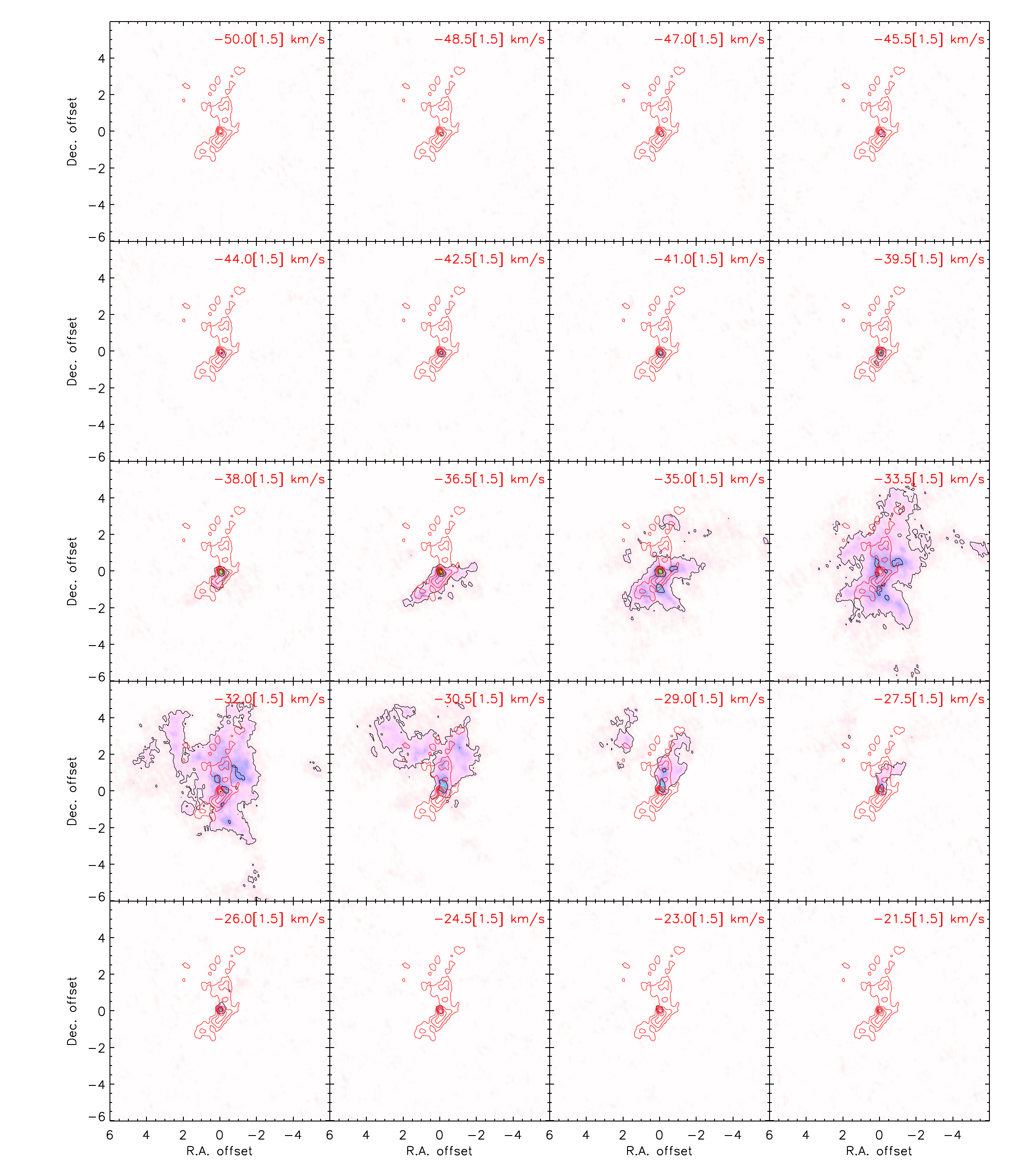}\\
\caption{Same as Figure \ref{fig:chanmap_12CO}, but for
the H$_2$S($2_{2,0} - 2_{1,1}$) emission. The black contours start at $5\sigma$
and have intervals of $10\sigma$ ($1\sigma=2.6~\mJybeam$).}
\label{fig:chanmap_H2S}
\end{center}
\end{figure}

Figures \ref{fig:chanmap_12CO} $-$ \ref{fig:chanmap_H2S} show the
channel maps of the $^{12}$CO ($2-1$), SiO ($5-4$), C$^{18}$O ($2-1$),
CH$_3$OH ($4_{2,2}-3_{1,2}$; E), H$_2$CO ($3_{2,1} - 2_{2,0}$), SO$_2$
($22_{2,20}-22_{1,21}$), and H$_2$S ($2_{2,0} - 2_{1,1}$) emissions.

\section{Moment Maps of the Molecular Lines on Large Scales}
\label{sec:appB}

\begin{figure}
\begin{center}
\includegraphics[width=0.9\textwidth]{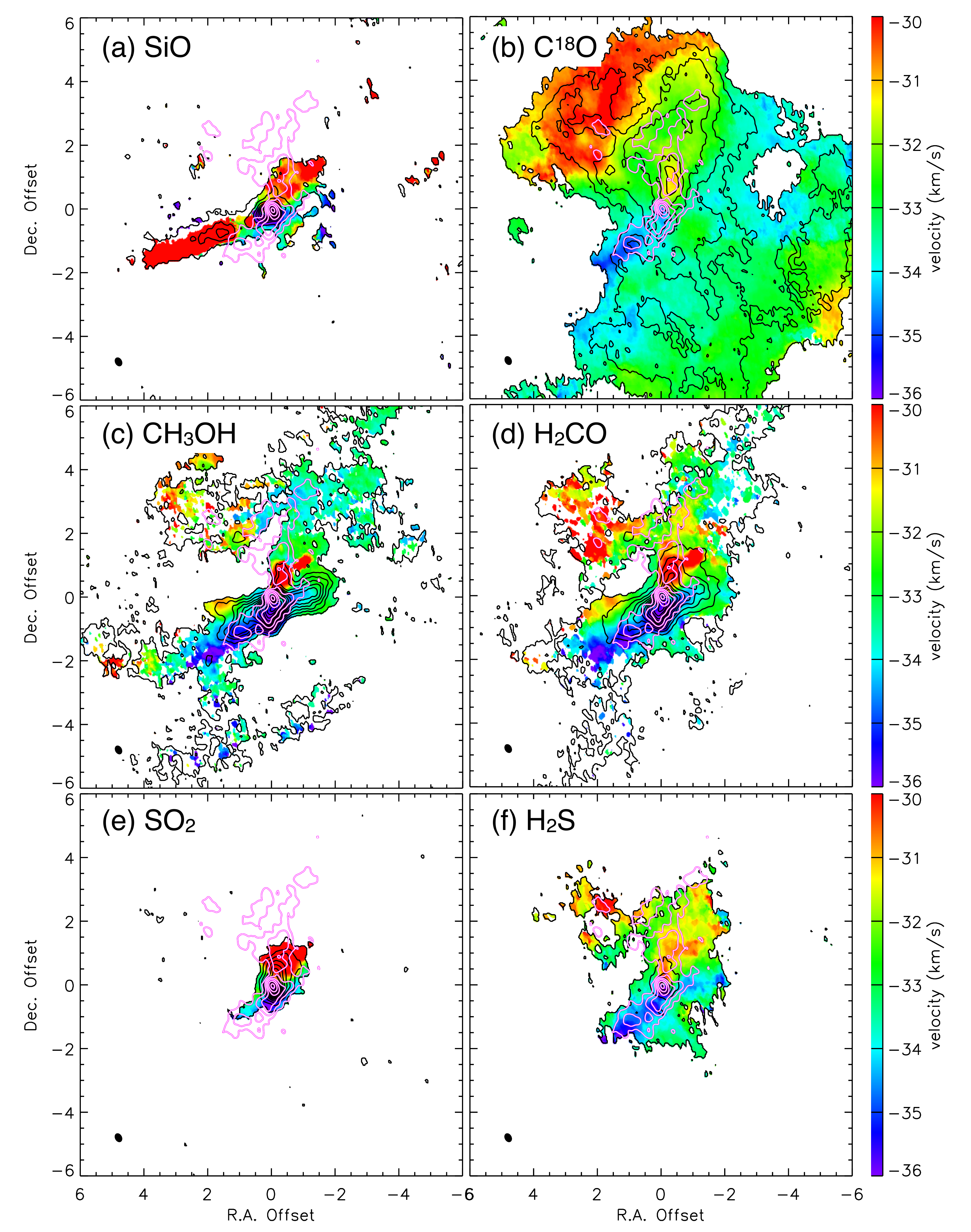}\\
\caption{Same as Figure \ref{fig:momentmap}, but showing the emissions on larger scales.}
\label{fig:momentmap_large}
\end{center}
\end{figure}

Figure \ref{fig:momentmap_large} shows the
moment 0 and 1 maps of the SiO ($5-4$), C$^{18}$O ($2-1$),
CH$_3$OH ($4_{2,2}-3_{1,2}$; E), H$_2$CO ($3_{2,1} - 2_{2,0}$), SO$_2$
($22_{2,20}-22_{1,21}$), and H$_2$S ($2_{2,0} - 2_{1,1}$) emissions on large scales.

\section{Kinematic Model Fitting of Rotating Disk without Radial Motion}
\label{sec:appC}

In \S\ref{sec:model}, we compared the observed PV diagrams of SiO,
CH$_3$OH, H$_2$CO, SO$_2$, and H$_2$S emissions with a model composed
of an infalling-rotating envelope on the outside and a Keplerian disk
on the inside. In such a model, the CH$_3$OH and H$_2$CO emissions, as
well as the low-velocity components of the SiO, SO$_2$, and H$_2$S
emissions, are explained by an infalling-rotating envelope. The
high-velocity components of the SiO emission are explained by a
Keplerian disk with its outer radius the same as the inner radius of the
envelope. The high-velocity components of the SO$_2$ and H$_2$S
emissions are explained by the same Keplerian disk, but truncated at an
inner boundary.  In this appendix, we explore the possibility that all
these molecular lines trace different parts of a single Keplerian
disk, i.e., with pure rotation but no infall motion.

\begin{figure}
\begin{center}
\includegraphics[width=0.7\textwidth]{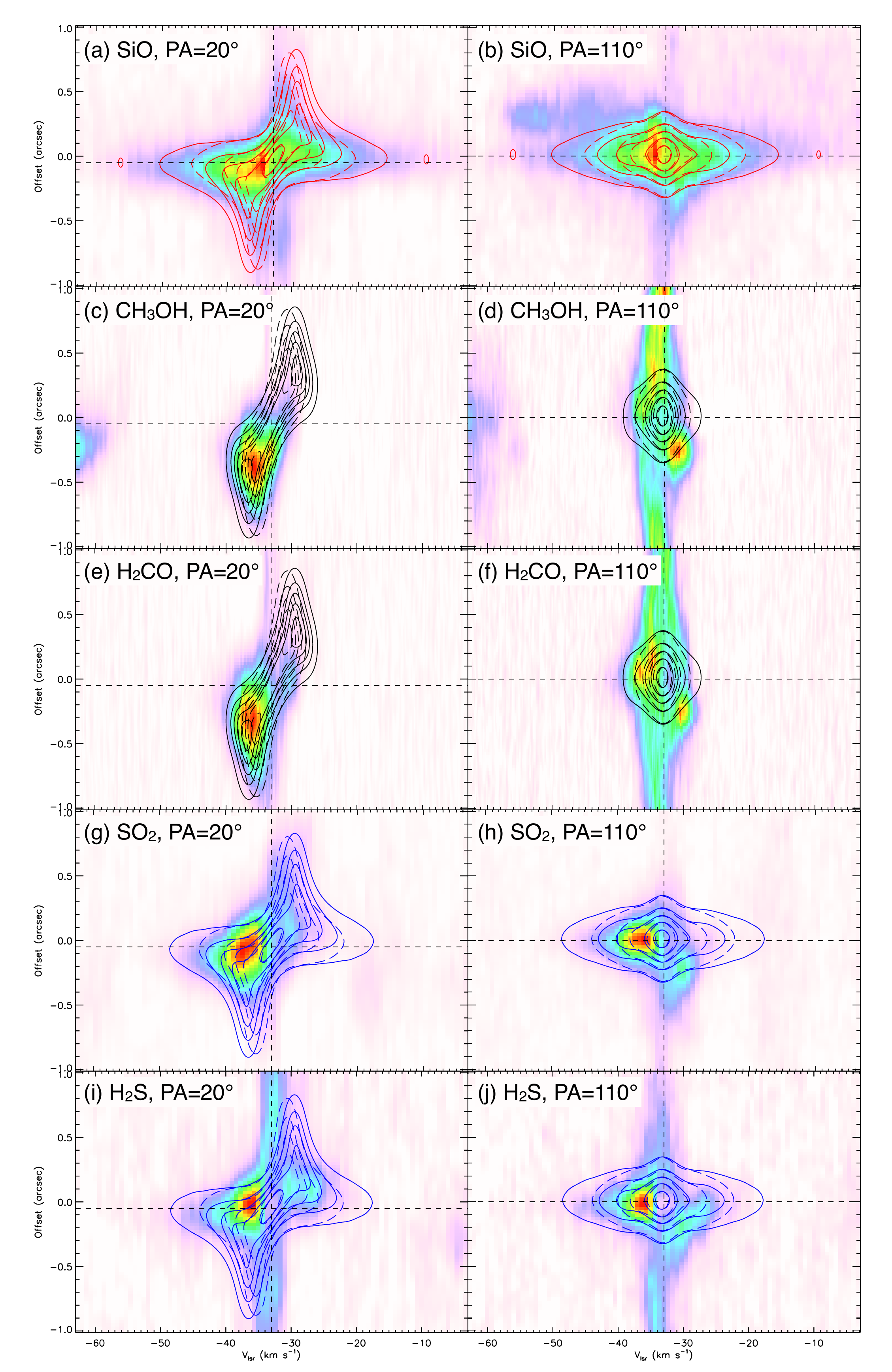}\\
\caption{
Same as Figure \ref{fig:model}, but showing models of only the Keplerian disk
with pure rotation but no radial motions. 
All the models (shown in contours) have the same outer boundary,
but different inner boundaries (see the text for details).
The solid contours show the best-fit disk model.
The dashed contours show the disk model in the best-fit disk$+$envelope model presented 
in \S\ref{sec:model} (Figure \ref{fig:model}) but extended to outer regions.
The left column shows the position-velocity diagrams along a cut perpendicular to the outflow axis 
(along the disk direction). The positive offsets are to the north
of the source. The right column shows the position-velocity diagrams
along a cut perpendicular to the disk direction. The positive
offsets are on the east side of the source.
The model contours are at levels of 0.1, 0.3, 0.5, 0.7, and 0.9 of the peak intensities.}
\label{fig:model_disk}
\end{center}
\end{figure}

Appendix Figure \ref{fig:model_disk} shows the comparison of such
models with the observations.  In these models, the Keplerian disk is
extended to an outer radius of $\rout/d=0.8\arcsec$ (same as the outer
radius of the envelope in the envelope$+$disk model presented in \S\ref{sec:model}).
The SiO PV diagrams are compared with models with the full disk, the
PV diagrams of the CH$_3$OH and H$_2$CO emissions are compared with
models of disks truncated at an inner radius of
$r_\mathrm{in}/d=0.25\arcsec$ (same as the inner radius of the
envelope in the envelope$+$disk model), and the PV
diagrams of the SO$_2$, and H$_2$S emissions are compared with models
with disks truncated at an inner radius of
$r_\mathrm{in}/d=0.02\arcsec$ (same as the inner radius of the
truncated disk model for these two lines presented in
\S\ref{sec:model}).  The disk height, density profile, and the
inclination are set to be same as those in the disk model in
\S\ref{sec:model}.  The only free parameter is the central
protostellar mass $m_*$, for which we explore within the range of 10
to 30 $M_\odot$ with an interval of $1~M_\odot$.  The best-fit model
shown in Appendix Figure \ref{fig:model_disk} (solid contours) has a
protostellar mass of $m_*=23~M_\odot$.  We also show another set of
models, in which the disk is same as that in the best-fit model
presented in \S\ref{sec:model} (i.e., around a protostar of
$m_*=11~M_\odot$; shown in dashed contours) for reference.

However, we do not consider that the pure disk models can explain the
observations as well as the disk$+$envelope model for several reasons.
First, the pure disk models cannot reproduce the observed velocity
gradients perpendicular to the disk direction (right column in
Appendix Figure \ref{fig:model_disk}), which are clearly seen in all
the molecular emissions.  Second, the velocity gradient along the disk
direction seen in the CH$_3$OH and H$_2$CO emissions (panels c and e)
are higher than that which a disk model can reproduce. For example,
the disk model with $m_*=23~M_\odot$ can reproduce well the emission
at offsets $<0.5\arcsec$ from the protostar, but has higher velocities
at offsets $>0.5\arcsec$. On the contrary, the disk model with
$m_*=11~M_\odot$ can reproduce the emission at offsets $>0.5\arcsec$
from the protostar, but underestimates the velocities at offsets
$<0.5\arcsec$. These suggest that, at least in the CH$_3$OH and
H$_2$CO emissions, the observed rotation profiles are steeper than
that of a Keplerian disk ($v_\varphi\propto r^{1/2}$) and more
consistent with that of an rotating-infalling envelope.  Third, the
pure-disk models cannot reproduce the red-shifted emissions in the
south and blue-shifted emission in the north, which are most clearly
seen in the SO$_2$, and H$_2$S emissions, but are also seen in the
other molecular lines.  Fourth, the best-fit pure disk model has a
central mass of $23~M_\odot$, from which a bolometric luminosity of
several $\times 10^5~L_\odot$ is expected (\citealt[]{ZT18}),
significantly higher than the observed luminosity (see
\S\ref{sec:target}).  Therefore, we consider that the transition from
an infalling-rotating envelope to a Keplerian disk is a better
explanation for the different spatial and kinematic patterns of the
molecular lines.

\clearpage
\renewcommand{\figurename}{Appendix Figure}

\clearpage

\clearpage

\clearpage

\clearpage

\clearpage

\clearpage

\clearpage

\clearpage

\end{document}